\documentclass[final]{ws-ijmpe}
\usepackage{graphicx}
\usepackage{ifpdf}
\usepackage{bm}
\usepackage{mathbbol}
\usepackage{mathrsfs}
\usepackage[makeroom]{cancel}
\def\bq{\bm{q}}
\def\bp{\bm{p}}
\def\bA{\bm{A}}
\def\bP{\bm{P}}
\def\bQ{\bm{Q}}
\def\bv{\bm{v}}
\def\bbeta{\bm{\beta}}
\def\rk{\mathrm{kin}}
\def\FC{F}
\def\e{\mathrm{e}}
\def\d{\mathrm{d}}
\def\Hv{e}
\def\HC{H^{\prime}}
\def\HCv{E}
\def\RB{\mathbb{R}}
\def\MB{\mathbb{M}}
\def\AB{\mathbb{A}}
\def\BB{\mathbb{B}}
\def\onehalf{{\textstyle\frac{1}{2}}}
\def\quarter{{\textstyle\frac{1}{4}}}
\def\pfrac#1#2{\frac{\partial #1}{\partial #2}}
\def\ppfrac#1#2#3{\frac{\partial^2 #1}{\partial #2\partial #3}}
\def\dfrac#1#2{\frac{\d #1}{\d #2}}

\begin{document}
\markboth{J.~Struckmeier}
{Extended Hamilton-Lagrange formalism}
%
%
\title{EXTENDED HAMILTON-LAGRANGE FORMALISM AND\\
ITS APPLICATION TO FEYNMAN'S PATH INTEGRAL\\
FOR RELATIVISTIC QUANTUM PHYSICS}
\author{J\"URGEN STRUCKMEIER}
\address{Gesellschaft f\"ur Schwerionenforschung (GSI),
Planckstrasse~1, 64291~Darmstadt, Germany\\
Goethe University, Max-von-Laue-Str.~1, 60438~Frankfurt am Main, Germany\\
struckmeier@fias.uni-frankfurt.de}
\maketitle
\begin{history}
\received{10 October 2008}
\end{history}
\begin{otherinfo}
Published in: Int.~J.~Mod.~Phys~E, Vol.~18, No.~1 (2009), pp.~79--108
\end{otherinfo}
\begin{abstract}
With this paper, a consistent and comprehensive treatise on the
foundations of the \emph{extended Hamilton-Lagrange formalism} will be presented.
In this formalism, the system's dynamics is parametrized along a
time-like system evolution parameter $s$, and the physical time
$t$ is treated as a \emph{dependent} variable $t(s)$ on equal
footing with all other configuration space variables $q^{i}(s)$.
In the action principle, the conventional classical action
$L\,dt$ is then replaced by the generalized action $L_{\e}ds$,
with $L$ and $L_{\e}$ denoting the conventional and the extended Lagrangian, respectively.
Supposing that both Lagrangians describe the same physical system
then provides the correlation of $L$ and $L_{\e}$.
In the existing literature, the discussion is restricted to only those
extended Lagrangians $L_{\e}$ that are \emph{homogeneous forms of first order} in the velocities.
As a result, the Legendre transformation of $L_{\e}$ to a
corresponding extended Hamiltonian is \emph{singular} and thus
does not provide us with an equivalent extended Hamiltonian $H_{\e}$.

In this paper, it is shown that a class of extended Lagrangians
$L_{\e}$ exists that are correlated to corresponding conventional
Lagrangians $L$ \emph{without being homogeneous functions in the velocities}.
Then the Legendre transformation of $L_{\e}$ to an \emph{extended Hamiltonian} $H_{\e}$ exists.
With this class of extended Hamiltonians, an extended canonical
formalism is presented that is completely analogous to the conventional Hamiltonian formalism.
The physical time $t$ and the negative \emph{value} of the conventional Hamiltonian
then constitute and an additional pair of conjugate canonical variables.
The extended formalism also includes a theory of extended canonical
transformations, where the time variable $t(s)$ is also subject to transformation.

In the extended formalism, the system's dynamics is described
as a motion on a hypersurface within an \emph{extended} phase space of even dimension.
It is shown that the hypersurface condition does \emph{not} embody
a constraint as the condition is automatically satisfied on the system path
that is given by the solution of the extended set of canonical equations.

It is furthermore demonstrated that the value of the extended Hamiltonian and the
parameter~$s$ constitute a second \emph{additional pair} of canonically conjugate variables.
In the corresponding quantum system, we thus encounter an additional \emph{uncertainty relation}.

As a consequence of the formal similarity of conventional and
extended Hamilton-Lagrange formalisms, Feynman's \emph{non-relativistic}
path integral approach can be converted \emph{on a general level}
into a form appropriate for \emph{relativistic} quantum physics.
In the emerging parametrized quantum description, the additional
uncertainty relation serves as the means to incorporate the
hypersurface condition and hence to finally eliminate the parametrization.

As the staring point, the non-homogeneous extended Lagrangian $L_{\e}$
of a classical relativistic point particle in an external electromagnetic field will be presented.
It will be shown that this extended Lagrangian can be transformed
into a corresponding extended Hamiltonian $H_{\e}$ by a \emph{regular} Legendre transformation.
With this $L_{\e}$, it is shown that the generalized path integral approach
yields the Klein-Gordon equation as the corresponding quantum description.
Moreover, the space-time propagator for a free relativistic particle will be derived.
These results can be regarded as the proof of principle of the
\emph{relativistic generalization} of Feynman's path integral approach to quantum physics.
\end{abstract}
\ccode{PACS numbers: 04.20.Fy, 03.65.-w, 03.65.Pm}
\section{Introduction}
Even more than hundred years after the emerging of Einstein's
special theory of relativity, the presentation of classical
dynamics in terms of the Lagrangian and the Hamiltonian formalisms
is still usually based in literature on the Newtonian absolute time as the system evolution
parameter\cite{abraham,greiner,arnold,frankel,saletan,ratiu,kleinert}.
The way to generalize the Hamilton-Lagrange formalism in order to render
it compatible with special relativity is obvious and well-established.
It consists of introducing a system evolution parameter, $s$,
as a new time-like independent variable, and of subsequently treating
the physical time $t=t(s)$ as a \emph{dependent} variable of $s$,
in parallel to all configuration space variables $q^{i}(s)$.
This idea has been pursued in numerous publications, only a few of them being cited here.

Despite this unambiguity in the foundations and the huge pile
of publications on the matter --- dating back to
P.~Dirac\cite{dirac} and C.~Lanczos\cite{lanczos} ---
a truly consistent extended Hamilton-Lagrange formalism is still missing.
The reason for this is that the discussion in the existing literature
is restricted to only those extended Lagrangians that are
\emph{homogeneous forms of first order} in the velocities.
In this paper, this class of Lagrangians will be referred to as \emph{trivial extended Lagrangians}.
For the class of trivial extended Lagrangians, corresponding trivial extended
Hamiltonians cannot be directly derived by a Legendre transformation as the transformation is singular.
Yet, trivial extended Hamiltonians can always be set of on the basis of a given conventional Hamiltonian.

As will be shown in this paper, \emph{extended} Lagrangians $L_{\e}$
indeed exist for given conventional Lagrangians $L$ that both describe
the same physical system and that are \emph{no homogeneous forms} in the velocities $\d q^{\mu}/\d s$.
In other words, the correlation of $L$ and $L_{\e}$ is \emph{not unique}
is the sense that we can find more than one extended Lagrangian $L_{\e}$
that can be reduced to the same conventional Lagrangian $L$.
This will be demonstrated for the simple case of the free relativistic point particle.

If for a given conventional Lagrangian $L$ a non-trivial extended Lagrangian
$L_{\e}$ can be found, then the Legendre transformation is regular, and hence an
equivalent extended Hamiltonian $H_{\e}\not\equiv0$ can be derived directly.
This will be shown for the case of a relativistic particle
in an external electromagnetic field, whose extended Hamiltonian
will be derived by Legendre-transforming the corresponding non-homogeneous extended Lagrangian.
Remarkably, we thus \emph{derive} an extended Hamiltonian which
coincides with the ``super-Hamiltonian'' that was \emph{postulated}
earlier by Misner, Thorne, and Wheeler\cite{misner}.

For extended Hamiltonians $H_{\e}$, the subsequent extended
set of canonical equations is found to perfectly coincide in its
\emph{form} with the conventional one.
This also applies for the theory of extended canonical transformations.
The \emph{trivial extended generating function}
$\FC_{2}$ is shown to generate exactly the subgroup of conventional
canonical transformations within the group of extended canonical transformations.
This subgroup consists of exactly those extended canonical mappings
that transform the time variables identically.

On grounds of the \emph{formal similarity} of conventional and extended
Hamilton-Lagrange formalisms, it is possible to formally convert
non-relativistic approaches that are based on conventional Lagrangians
into relativistic approaches in terms of extended Lagrangians.
This idea is worked out exemplarily for Feynman's path integral
approach to quantum physics.\cite{feynman}

The paper is organized as follows.
We start in Sect.~\ref{sec:eleq} with the Lagrangian description
and derive from the extended form of the action integral the
extended Lagrangian $L_{\e}$, together with its relation to the conventional Lagrangian $L$.
It is shown that this relation reduces to the factor $\d t/\d s$.
The extended set of Euler-Lagrange equations then follows
from the dependencies of the extended Lagrangian.

In the extended Hamilton-Lagrange description of dynamics, the system's
motion takes place on \emph{hypersurfaces in extended phase spaces}.
In the extended Lagrangian formalism, this space is given by the tangent
bundle $T(\MB\times\RB)$, whereas in the extended Hamiltonian formalism,
the hypersurface lies within the cotangent bundle $T^{*}(\MB\times\RB)$,
both cases built over the space-time configuration manifold $\MB\times\RB$.
It is proved that the emerging of a hypersurface condition does
\emph{not} imply the system to be constrained as the condition
is always satisfied on the system path that is given by the solution
of the (unconstrained) extended set of canonical equations.
This perception corresponds to the case of a conventional Hamiltonian
system with no explicit time dependence, where the system's motion
takes place on a phase-space hypersurface of constant energy.
Likewise, the correlation of the dynamical variables that is induced
by this hypersurface of constant energy is not considered to be a
constraint as for autonomous systems the energy is automatically
maintained by any solution of the set of canonical equations.
The hypersurface condition thus distinguishes physical from unphysical
phase-space locations that cannot represent at any time the system's
state for the given canonical equations and the initial conditions.
In this sense, the hypersurface condition is the classical particle
analogue of the \emph{mass shell condition} of quantum field theory.

To provide a simple example, we derive in Sect.~\ref{sec:lag1-fp} the
non-homogeneous extended Lagrangian $L_{\e}$ for a free relativistic point particle.
This Lorentz-invariant Lagrangian $L_{\e}$ has the remarkable
feature to be \emph{quadratic} in the velocities.
This contrasts with the conventional Lorentz-invariant
Lagrangian $L$ that describes the identical dynamics.
For this system, the hypersurface condition depicts
the constant square of the four-velocity vector.

We show in Sect.~\ref{sec:lag1-em} that the extended
Lagrangian $L_{\e}$ of a relativistic particle in an external
electromagnetic field agrees \emph{in its form} with the
corresponding non-relativistic conventional Lagrangian $L$.
The difference between both is that the derivatives in the
extended Lagrangian $L_{\e}$ are being defined with respect
to the particle's \emph{proper time}, which are converted into
derivatives with respect to the Newtonian \emph{absolute time} in the non-relativistic limit.

In Sect.~\ref{sec:caneq}, we switch to the extended
Hamiltonian description.
As the extended Hamiltonian $H_{\e}$ springs up from a non-homogeneous
extended Lagrangian $L_{\e}$ by means of a regular Legendre
transformation, both functions equally contain the total
information on the dynamical system in question.
The Hamiltonian counterparts of the Lagrangian description,
namely, the extended set of canonical equations, the hypersurface
condition, and the correlation of the extended Hamiltonian $H_{\e}$
to the conventional Hamiltonian $H$ are presented.
On this basis, the theory of extended canonical
transformations and the extended version of the
Hamilton-Jacobi equation are worked out as
straightforward generalizations of the conventional theory.
As a mapping of the time $t$ is incorporated in an extended
canonical transformation, not only the transformed coordinates
emerging from the Hamilton-Jacobi equation are constants,
as usual, but also the transformed time $T$.
The extended Hamilton-Jacobi equation may thus be interpreted
as defining the mapping of the entire dynamical system into
its state at a \emph{fixed instant of time}, i.e., for instance,
into its initial state.
In the extended formulation, the Hamilton-Jacobi
equation thus reappears in a new perspective.

We furthermore show that the \emph{value} of the extended
Hamiltonian $H_{\e}$ and the system evolution parameter $s$
yield an additional pair of canonically conjugate variables.
For the corresponding quantum system, we thus encounter
an additional \emph{uncertainty relation}.
Based on both the extended Lagrangian $L_{\e}$ and the additional
uncertainty relation, we present in Sect.~\ref{sec:pathint}
the path integral formalism in a form appropriate for
\emph{relativistic quantum systems}.
An extension of Feynman's approach was worked out
earlier\cite{duru} for a particular system.
Nevertheless, the most general form of the extended path
integral formalism that applies for any extended Lagrangian
$L_{\e}$ is presented here for the first time.
By consistently treating space and time variables on
equal footing, the generalized path integral formalism
is shown to apply as well for Lagrangians that
\emph{explicitly} depend on time.
In particular, the transition of a wave function is presented
here as a space-time integral over a space-time propagator.
In this context, we address the physical meaning of the
additional integration over $t$.
The uncertainty relation is exhibited as the \emph{quantum
physics' means} to incorporate the hypersurface condition
in order to finally eliminate the parameterization.

On grounds of a generalized understanding of the action principle,
Feynman showed that the Schr\"odinger equation emerges as the
non-relativistic \emph{quantum description} of a dynamical system if the
corresponding \emph{classical} system is described by the non-relativistic
Lagrangian $L$ of a point particle in an external potential.
Parallel to this beautiful approach, we derive in
Section~\ref{sec:kg} the Klein-Gordon equation as the
\emph{relativistic quantum description} of a system,
whose classical counterpart is described by the \emph{non-homogeneous
extended Lagrangian} $L_{\e}$ of a relativistic
point particle in an external electromagnetic field.
The reason for this to work is twofold.
As the extended Lagrangian $L_{\e}$ agrees in its
form with the conventional non-relativistic Lagrangian
$L$, the generalized path integral formalism can be
worked out similarly to the non-relativistic case.
Furthermore, as we proceed in our derivation an
infinitesimal proper time step $\Delta s$ only and consider
the limit $\Delta s\to0$, the hypersurface condition disappears
by virtue of the uncertainty relation.

We finally derive in Sect.~\ref{sec:prop} the space-time
propagator for the wave function of a free particle with
spin zero from the extended Lagrangian of a free
relativistic point particle.
The hypersurface condition, as the companion of the classical
extended description, is taken into account in the quantum
description by integrating over all possible
parameterizations of the system's variables.
This integration is now explained in terms of the
uncertainty relation.
We regard these results as the ultimate confirmation of the
relativistic generalization of Feynman's path integral formalism.
\section{Extended Hamilton-Lagrange formalism}
\subsection{\label{sec:eleq}Extended set of Euler-Lagrange equations}
The conventional formulation of the principle of least action
is based on the action functional $S[\bq(t)]$, defined by
\begin{equation}\label{principle0}
S[\bq(t)]=\int_{t_{a}}^{t_{b}}L\left(\bq,\dfrac{\bq}{t},t\right)\d t,
\end{equation}
with $L(\bq,\dot{\bq},t)$ denoting the system's conventional Lagrangian, and the vector
of configuration space variables $\bq(t)=(q^{1}(t),\ldots,q^{n}(t))$ as a function of time.
In this formulation, the independent variable time $t$ plays
the role of the Newtonian \emph{absolute time}.
The actual path $(\bar{\bq}(t),\dot{\bar{\bq}}(t))$ the physical system ``realizes''
is given as the \emph{extremum} of the action $S$, hence for $\delta S=0$.
The path representing this extremum of $S$ is the solution of the
set of Euler-Lagrange equations ($i=1,\ldots,n$) for the given
initial conditions $\bq_{0},\dot{\bq}_{0}$,
\begin{equation}\label{lageqm0}
\dfrac{}{t}\left(\pfrac{L}{\left(\dfrac{q^{i}}{t}
\right)}\right)-\pfrac{L}{q^{i}}=0.
\end{equation}
The reformulation of the least action principle~(\ref{principle0})
that is eligible for relativistic physics is accomplished by treating
the time $t(s)=q^{0}(s)/c\,$ --- like the vector $\bq(s)$ of configuration
space variables --- as a \emph{dependent} variable of a newly
introduced timelike independent variable, $s$~\cite{lanczos,fanchi,rohrlich,struck}.
The action functional then writes in terms of an
\emph{extended Lagrangian} $L_{\e}$
\begin{equation}\label{principle1}
S_{\e}[\bq(s),t(s)]=\int_{s_{a}}^{s_{b}}L_{\e}\left(\bq,\dfrac{\bq}{s},t,
\dfrac{t}{s}\right)\d s\equiv
\int_{s_{a}}^{s_{b}}L_{\e}\left(q^{\mu},\dfrac{q^{\mu}}{s}\right)\d s.
\end{equation}
Herein, the index $\mu=0,\ldots,n$ denotes the entire range of
extended configuration space variables.
As the action functional~(\ref{principle1}) has the form of
(\ref{principle0}), the subsequent Euler-Lagrange equations that determine
the particular path $(\bar{\bq}(s),\bar{t}(s))$ on which the value of the
functional~(\ref{principle1}) takes on an extreme value,
adopt the customary form of Eq.~(\ref{lageqm0})
\begin{equation}\label{lageqm}
\dfrac{}{s}\left(\pfrac{L_{\e}}{\left(\dfrac{q^{\mu}}{s}
\right)}\right)-\pfrac{L_{\e}}{q^{\mu}}=0.
\end{equation}
For the index $\mu=0$, the Euler-Lagrange equation can be
expressed equivalently in terms of $t(s)$ as
\begin{equation}\label{lageqm-t}
\dfrac{}{s}\left(\pfrac{L_{\e}}{\left(\dfrac{t}{s}
\right)}\right)-\pfrac{L_{\e}}{t}=0.
\end{equation}
The equations of motion for both $\bq(s)$ and $t(s)$
are thus determined by the extended Lagrangian $L_{\e}$.
The solution $\bq(t)$ of the Euler-Lagrange equations that equivalently
emerges from the corresponding conventional Lagrangian $L$ may then
be constructed by eliminating the evolution parameter $s$.

As the actions, $S$ and $S_{\e}$, are supposed to be alternative
characterizations of the \emph{same} underlying physical system,
the action principles $\delta S=0$ and $\delta S_{\e}=0$ must
hold simultaneously.
This means that
$$
\delta\int_{s_{a}}^{s_{b}}L\dfrac{t}{s}\,\d s=
\delta\int_{s_{a}}^{s_{b}}L_{\e}\,\d s,
$$
which, in turn, is assured if both integrands differ at most by
the $s$-derivative of an arbitrary differentiable function $\FC(\bq,t)$
$$
L\dfrac{t}{s}=L_{\e}+\dfrac{\FC}{s}.
$$
Functions $\FC(\bq,t)$ define a particular class of point
transformations of the dynamical variables, namely those ones
that preserve the form of the Euler-Lagrange equations.
Such a transformation can be applied at any time in the
discussion of a given Lagrangian system and should be
distinguished from correlating $L_{\e}$ and $L$.
We may thus restrict ourselves without loss of generality
to those correlations of $L$ and $L_{\e}$, where $\FC\equiv0$.
In other words, we correlate $L$ and $L_{\e}$ \emph{without}
performing simultaneously a transformation of the dynamical variables.
We will discuss this issue in the more general context of
\emph{extended canonical transformations} in Sect.~\ref{sec:cantra}.
The extended Lagrangian $L_{\e}$ is then related to
the conventional Lagrangian, $L$, by
\begin{equation}\label{lag1}
L_{\e}\left(\bq,\dfrac{\bq}{s},t,\dfrac{t}{s}\right)=
L\left(\bq,\dfrac{\bq}{t},t\right)\dfrac{t}{s},\qquad
\dfrac{\bq}{t}=\dfrac{\bq/\d s}{t/\d s}.
\end{equation}
The derivatives of $L_{\e}$ from Eq.~(\ref{lag1}) with respect
to its arguments can now be expressed in terms of the conventional
Lagrangian $L$ as
\begin{align}
\quad\pfrac{L_{\e}}{q^{\mu}}&=\pfrac{L}{q^{\mu}}\dfrac{t}{s},
\qquad\mu=1,\ldots,n\\
\quad\pfrac{L_{\e}}{t}&=\pfrac{L}{t}
\dfrac{t}{s}\\
\pfrac{L_{\e}}{\left(\dfrac{q^{\mu}}{s}\right)}&=
\pfrac{L}{\left(\dfrac{q^{\mu}}{t}\right)},\qquad
\mu=1,\ldots,n\label{L1-deri}\\
\pfrac{L_{\e}}{\left(\dfrac{t}{s}\right)}&=L+
\sum_{\mu=1}^{n}\pfrac{L}{\left(\dfrac{q^{\mu}}{t}\right)}
\pfrac{\left(\dfrac{q^{\mu}/\d s}{t/\d s}\right)}{\left(\dfrac{t}{s}\right)}\dfrac{t}{s}=
L-\sum_{\mu=1}^{n}\pfrac{L}{\left(\dfrac{q^{\mu}}{t}\right)}\dfrac{q^{\mu}}{s}
{\left(\dfrac{s}{t}\right)}^{2}\dfrac{t}{s}\nonumber\\
&=L-\sum_{\mu=1}^{n}\pfrac{L}{\left(\dfrac{q^{\mu}}{t}\right)}
\dfrac{q^{\mu}}{t}.\label{L1-deri2}
\end{align}
Equations~(\ref{L1-deri}) and (\ref{L1-deri2}) yield for the
following sum over the extended range $\mu=0,\ldots,n$ of
dynamical variables
\begin{align*}
\sum_{\mu=0}^{n}\pfrac{L_{\e}}{\left(\dfrac{q^{\mu}}{s}\right)}
\dfrac{q^{\mu}}{s}&=L\dfrac{t}{s}-
\sum_{\mu=1}^{n}\pfrac{L}{\left(\dfrac{q^{\mu}}{t}\right)}
\dfrac{q^{\mu}}{t}\dfrac{t}{s}+
\sum_{\mu=1}^{n}\pfrac{L}{\left(\dfrac{q^{\mu}}{t}\right)}
\dfrac{q^{\mu}}{s}\\
&=L_{\e}.
\end{align*}
The extended Lagrangian $L_{\e}$ thus satisfies the equation
\begin{equation}\label{lagid}
L_{\e}-\sum_{\mu=0}^{n}\pfrac{L_{\e}}{\left(\dfrac{q^{\mu}}{s}\right)}
\dfrac{q^{\mu}}{s}
\begin{cases}\stackrel{\not\equiv}{=}0 & \mbox{ if $L_{\e}$ not homogeneous}\\
\equiv0 & \mbox{ if $L_{\e}$ homogeneous}\end{cases}
\mbox{ in }\dfrac{q^{\mu}}{s}.
\end{equation}
Regarding the correlation~(\ref{lag1}) and the pertaining
condition~(\ref{lagid}), two different cases must be distinguished.
In the first case, an extended Lagrangian $L_{\e}$ can be set up immediately
by multiplying a given conventional Lagrangian $L$ with $\d t/\d s$
and expressing all velocities $\d\bq/\d t$ in terms of $\d\bq/\d s$
according to Eq.~(\ref{lag1}).
Such an extended Lagrangian $L_{\e}$ is called a \emph{trivial extended Lagrangian}
as it contains no additional information on the underlying dynamical system.
A trivial extended Lagrangian $L_{\e}$ constitutes a homogeneous form of first order
in the $n+1$ variables $\d q^{0}/\d s,\ldots,\d q^{n}/\d s$.
This may be seen by replacing all derivatives $\d q^{\mu}/\d s$
with $a\cdot\d q^{\mu}/\d s$, $a\in\RB$ in Eq.~(\ref{lag1}), which yields
\begin{align*}
L_{\e}\left(\bq,a\dfrac{\bq}{s},t,a\dfrac{t}{s}\right)&=
L\left(\bq,\dfrac{\bq}{t},t\right)a\dfrac{t}{s}\\
&=aL_{\e}\left(\bq,\dfrac{\bq}{s},t,\dfrac{t}{s}\right).
\end{align*}
Consequently, Euler's theorem on homogeneous functions
states that Eq.~(\ref{lagid}) constitutes an
\emph{identity}\cite{lanczos}.
In that case, we may differentiate the identity with respect to
the velocity $\d q^{\nu}/\d s$ to get
\begin{equation}\label{lagid-deri}
\sum_{\mu=0}^{n}\pfrac{^{2}L_{\e}}
{\left(\dfrac{q^{\mu}}{s}\right)\partial\left(\dfrac{q_{\nu}}{s}\right)}\dfrac{q^{\mu}}{s}\equiv0.
\end{equation}
This is a homogeneous set of $n$ equations for the velocities $\d q^{\mu}/\d s$.
It has a non-trivial solution ($\d\bq/\d s\neq0$)
only if the coefficient matrix is singular
\begin{equation}\label{singularLegTrans}
\det\left(\pfrac{^{2}L_{\e}}
{\left(\dfrac{q^{\mu}}{s}\right)\partial\left(\dfrac{q_{\nu}}{s}\right)}\right)=0.
\end{equation}
Due to Eq.~(\ref{singularLegTrans}), a corresponding extended Hamiltonian
$H_{\e}$ does not follow from a trivial extended Lagrangian $L_{\e}$
as the mediating Legendre transformation is \emph{singular}.

The Euler-Lagrange equation~(\ref{lageqm-t}) for $\d t/\d s$ then
reduces to the conventional set of Eqs.~(\ref{lageqm0}) for arbitrary
$t(s)$, hence, we do not obtain a substantial equation of motion for $t(s)$.
Inserting Eq.~(\ref{L1-deri2}) into Eq.~(\ref{lageqm-t}), one finds
$$
\sum_{\mu=1}^{n}\underbrace{\dfrac{q^{\mu}}{t}}_{\neq0}\bigg[
\underbrace{\pfrac{L}{q^{\mu}}-\dfrac{}{t}\left(
\pfrac{L}{\dot{q}^{\mu}}\right)}_{\Rightarrow\;\;=0}\bigg]=0.
$$
The parametrization of time $t(s)$ is thus left undetermined ---
which reflects the fact that a conventional Lagrangian does not
provide any information on a parametrization of time and that
a trivial extended Lagrangian does not incorporate additional information.

The second case is completely overlooked in literature (cf, for instance\cite{dirac,lanczos,johns}), namely that extended Lagrangians
$L_{\e}$ exist that are related to a given conventional Lagrangian $L$
according to Eq.~(\ref{lag1}) \emph{without being homogeneous forms}
in the $n+1$ velocities $\d q^{\mu}/\d s$.
In Sect.~\ref{sec:lag1-fp}, a simple example will be furnished by setting up
such a non-homogeneous extended Lagrangian $L_{\e}$ for the \emph{free relativistic point particle}.
For a non-homogeneous extended Lagrangian $L_{\e}$, the extended set of
Euler-Lagrange equations~(\ref{lageqm}) is not redundant
and the Legendre transformation to an extended Hamiltonian $H_{\e}$ exists.
In that case, Eq.~(\ref{lagid}) does \emph{not} represent
an identity, which implies that Eq.~(\ref{lagid-deri})
and, subsequently, Eq.~(\ref{singularLegTrans}) do not hold.
Then, Eq.~(\ref{lagid}), regarded as an \emph{implicit equation},
is always satisfied on the extended system evolution
path parametrized by $s$, which is given by the solution of
the extended set of Euler-Lagrange equations~(\ref{lageqm}).
This can be seen by calculating the total $s$-derivative of Eq.~(\ref{lagid})
and inserting the Euler-Lagrange equations~(\ref{lageqm})
\begin{align}
&\quad\,\dfrac{}{s}L_{\e}\left(q^{\mu},\dfrac{q^{\mu}}{s}\right)-
\sum_{\mu=0}^{n}\dfrac{q^{\mu}}{s}\dfrac{}{s}\pfrac{L_{\e}}
{\left(\dfrac{q^{\mu}}{s}\right)}-\sum_{\mu=0}^{n}\pfrac{L_{\e}}
{\left(\dfrac{q^{\mu}}{s}\right)}\dfrac{\left(\dfrac{q^{\mu}}{s}\right)}{s}\nonumber\\
&=\dfrac{L_{\e}}{s}-\sum_{\mu=0}^{n}\pfrac{L_{\e}}{q^{\mu}}\dfrac{q^{\mu}}{s}-\sum_{\mu=0}^{n}
\pfrac{L_{\e}}{\left(\dfrac{q^{\mu}}{s}\right)}\dfrac{\left(\dfrac{q^{\mu}}{s}\right)}{s}\nonumber\\
&=0.
\label{lagid-deri2}
\end{align}
For this reason, Eq.~(\ref{lagid}) actually does \emph{not} impose a constraint
on the system's evolution along~$s$ but separates \emph{unphysical states}
that do not satisfy Eq.~(\ref{lagid}) from the physical states that are
solutions of the Euler-Lagrange equations~(\ref{lageqm}).
In this respect, Eq.~(\ref{lagid}) exactly corresponds to the case of the conserved energy function
$\Hv(t)=\Hv_{0}$ of a conventional Lagrangian system $L(\bq,\dot{\bq})$ with no explicit time dependence.
In that case, the quantity $\Hv(t)$
\begin{equation}\label{lagid-conv}
\Hv(t)=\sum_{\mu=1}^{n}\pfrac{L}{\dot{q}^{\mu}}\dot{q}^{\mu}-L(\bq,\dot{\bq})=\Hv_{0}
\end{equation}
is a constant of motion and hence defines a surface in $TM$ on which the system's
motion takes place.
Nevertheless, it is not considered a constraint as the condition~(\ref{lagid-conv}) is
automatically satisfied by means of the conventional Euler-Lagrange equations~(\ref{lageqm0}),
\begin{align*}
\dfrac{\Hv(t)}{t}&=\sum_{\mu=1}^{n}\left(
\dot{q}^{\mu}\dfrac{}{t}\pfrac{L}{\dot{q}^{\mu}}+
\cancel{\pfrac{L}{\dot{q}^{\mu}}\ddot{q}^{\mu}}-
\pfrac{L}{q^{\mu}}\dot{q}^{\mu}-\cancel{\pfrac{L}{\dot{q}^{\mu}}\ddot{q}^{\mu}}\right)\\
&=\sum_{\mu=1}^{n}\dot{q}^{\mu}\underbrace{\left(\dfrac{}{t}\pfrac{L}{\dot{q}^{\mu}}-\pfrac{L}{q^{\mu}}\right)}_{=0}\\
&=0.
\end{align*}
To summarize, by switching from the conventional variational
principle~(\ref{principle0}) to the extended
representation~(\ref{principle1}), we have introduced an extended
Lagrangian $L_{\e}$ that in addition depends on $\d t(s)/\d s$.
Due to the emerging conserved quantity that follows from Eq.~(\ref{lagid-deri2}),
the actual number of degrees of freedom is unchanged.
In the language of Differential Geometry, the system's motion along the parameter $s$ now takes place
on a \emph{hypersurface}, defined by Eq.~(\ref{lagid}), within the tangent bundle $T\MB_\e\equiv T(\MB\times\RB)$
over the ``spatial-plus-time'' configuration manifold \mbox{$\MB_\e\equiv\MB\times\RB$}.
This is the basis required for the description of \emph{relativistic} point particle dynamics,
which mandates configuration space coordinates and time to be treated on equal footing in a chart representation.
It contrasts with the conventional Lagrangian description in $T\MB$ over the spatial
configuration manifold $\MB$ for Lagrangians that do not explicitly depend on the system's parameter $t$,
which is commonly identified in applications with Newton's absolute time.
\subsection{\label{sec:caneq}Extended set of canonical equations}
The Lagrangian formulation of particle dynamics can
\emph{equivalently} be expressed as a Hamiltonian description.
The complete information on the given dynamical system is then
contained in a Hamiltonian $H$, which carries the same
information content as the corresponding Lagrangian $L$.
It is defined by the Legendre transformation
\begin{equation}\label{legendre}
H(\bq,\bp,t)=\sum_{\mu=1}^{n}p_{\mu}\dfrac{q^{\mu}}{t}-
L\left(\bq,\dfrac{\bq}{t},t\right),
\end{equation}
with the covariant momentum vector components $p_{\mu}$ being defined by
$$
p_{\mu}=\pfrac{L}{\left(\dfrac{q^{\mu}}{t}\right)}.
$$
Correspondingly, the \emph{extended} Hamiltonian $H_{\e}$ is defined
as the extended Legendre transform of the extended Lagrangian $L_{\e}$ as

\begin{equation}\label{legendre1}
H_{\e}(\bq,\bp,q^{0},p_{0})=\sum_{\mu=0}^{n}p_{\mu}\dfrac{q^{\mu}}{s}-
L_{\e}\left(q^{\nu},\dfrac{q^{\nu}}{s}\right),
\end{equation}
wherein $q^{0}(s)=ct(s)$ and $p_{0}(s)$ denotes the canonical conjugate variable of $q^{0}(s)$.
In order for $H_{\e}$ to take over the complete information on the
dynamical system from $L_{\e}$, the Hesse matrix must be non-singular
$$
\det\left(\pfrac{^{2}L_{\e}}{\left(\dfrac{q^{\mu}}{s}\right)
\partial\left(\dfrac{q_{\nu}}{s}\right)}\right)\ne0.
$$
We know from Eq.~(\ref{L1-deri}) that for $\mu=1,\ldots,n$
the momentum variable $p_{\mu}$ is equally obtained from
the extended Lagrangian $L_{\e}$,
\begin{equation}\label{p-def}
p_{\mu}=\pfrac{L_{\e}}{\left(\dfrac{q^{\mu}}{s}\right)}.
\end{equation}
This fact ensures the Legendre transformations~(\ref{legendre})
and (\ref{legendre1}) to be compatible.
For the index $\mu=0$, i.e., for $q^{0}=ct$ we must take some care
as the derivative of $L_{\e}$ with respect to $\d t/\d s$ evaluates to
$$
\pfrac{L_{\e}}{\left(\dfrac{t}{s}\right)}=L-\sum_{\mu=1}^{n}
\pfrac{L}{\left(\dfrac{q^{\mu}}{t}\right)}\dfrac{q^{\mu}}{t}=-H(\bq,\bp,t).
$$
The momentum coordinate $p_{0}(s)$ that is conjugate to $q^{0}=ct(s)$
must therefore be defined as
\begin{equation}\label{p0-def0}
p_{0}(s)=-\frac{\Hv(s)}{c},\qquad\Hv(s)\stackrel{\not\equiv}{=}
H\big(\bq(s),\bp(s),t(s)\big),
\end{equation}
with $\Hv(s)$ representing the instantaneous \emph{value} of the
Hamiltonian $H$ at $s$, but \emph{not} the \emph{function} $H$
proper as these functions are different.
The canonical coordinate $p_{0}$ must be conceived --- like all
other canonical coordinates --- as a function of the independent
variable, $s$, only.
Thus, $p_{0}$ has solely a derivative with respect to $s$.
In contrast, the Hamiltonian $H$ contains the complete information
on the underlying dynamical system --- which is provided as the
dependence of the value $\Hv(s)$ of $H$ on the \emph{individual}
values of the $q^{\mu}(s)$, $p_{\mu}(s)$, and $t(s)$ --- and thus
has derivatives with respect to all these canonical coordinates.
We may express the definition of $p_{0}(s)$, and $\Hv(s)$,
by means of the comprehensible notation
\begin{equation}\label{p0-def}
p_{0}(s)=\pfrac{L_{\e}}{\left(\dfrac{q^{0}}{s}\right)}(s)
\quad\Longleftrightarrow\quad
\Hv(s)=-\pfrac{L_{\e}}{\left(\dfrac{t}{s}\right)}(s).
\end{equation}
According to the extended Legendre transformation~(\ref{legendre1}),
the condition~(\ref{lagid}) translates in the
extended Hamiltonian description simply into
\begin{equation}\label{hamid}
H_{\e}\big(\bq(s),\bp(s),t(s),\Hv(s)\big)=0.
\end{equation}
This means that the extended Hamiltonian $H_{\e}$ directly
defines the hypersurface within the extended phase space
the classical particle motion is restricted to.
Geometrically, the hypersurface lies in the cotangent bundle
$T^*\MB_\e\equiv T^{*}(\MB\times\RB)$ over the same extended configuration
manifold $\MB_\e\equiv\MB\times\RB$ as in the case of the Lagrangian description.
This is exactly the higher-dimensional analogue of the
case of an \emph{autonomous} conventional Hamiltonian system,
hence a Hamiltonian with no \emph{explicit} time dependence,
$H\big(\bq(t),\bp(t)\big)=\Hv_{0}$ --- where the system's initial
energy $\Hv_{0}$ embodies a \emph{constant of motion}.
In that case, the system's motion again takes place on a
hypersurface that is now defined by $H(\bq,\bp)=\Hv_{0}$ and
represents the phase-space surface of constant energy within the
cotangent bundle $T^{*}\MB$ over the configuration manifold $\MB$.

By virtue of the Legendre transformations~(\ref{legendre})
and~(\ref{legendre1}), the correlation from Eq.~(\ref{lag1})
of extended and conventional Lagrangians is finally converted into
\begin{align}
H_{\e}(\bq,\bp,t,\Hv)&=\sum_{\mu=1}^{n}p_{\mu}\dfrac{q^{\mu}}{s}-
\Hv\dfrac{t}{s}-L_{\e}\left(\bq,\dfrac{\bq}{s},t,\dfrac{t}{s}\right)\nonumber\\
&=\sum_{\mu=1}^{n}p_{\mu}\dfrac{q^{\mu}}{s}-
\Hv\dfrac{t}{s}-L\left(\bq,\dfrac{\bq}{s},t\right)\dfrac{t}{s}\nonumber\\
&=\cancel{\sum_{\mu=1}^{n}p_{\mu}\dfrac{q^{\mu}}{s}}-\Hv\dfrac{t}{s}+
\left(H(\bq,\bp,t)-\cancel{\sum_{\mu=1}^{n}p_{\mu}\dfrac{q^{\mu}}{t}}\right)\dfrac{t}{s}\nonumber\\
&=\big(H(\bq,\bp,t)-\Hv\big)\dfrac{t}{s}.
\label{H1-def}
\end{align}
The extended Legendre transformation~(\ref{legendre1})
in conjunction with (\ref{p-def}) and the extended set of
Euler-Lagrange equations~(\ref{lageqm}) immediately
yields the extended set of canonical equations ($\mu=0,\ldots,n$),
\begin{equation}\label{caneq-def}
\pfrac{H_{\e}}{p_{\mu}}=\dfrac{q^{\mu}}{s},\qquad
\pfrac{H_{\e}}{q^{\mu}}=-\pfrac{L_{\e}}{q^{\mu}}=-\dfrac{p_{\mu}}{s}.
\end{equation}
The right-hand sides of these equations follow directly from the
Legendre transformation~(\ref{legendre1}) as the Lagrangian
$L_{\e}$ does not depend on the momenta $p_{\mu}$ and has, up to
the sign, the same space-time dependence as the Hamiltonian $H_{\e}$.
The extended set is characterized by the additional pair of
canonical equations for the index $\mu=0$, which reads in
terms of $t(s)$ and $\Hv(s)$
\begin{equation}\label{caneq-def1}
\dfrac{\Hv}{s}=\pfrac{H_{\e}}{t},\qquad
\dfrac{t}{s}=-\pfrac{H_{\e}}{\Hv}.
\end{equation}
For the total derivative of $H_{\e}(\bq,\bp,t,\Hv)$ we thus find
\begin{align*}
\dfrac{H_{\e}}{s}&=\pfrac{H_{\e}}{p_{i}}\dfrac{p_{i}}{s}+
\pfrac{H_{\e}}{q^{i}}\dfrac{q^{i}}{s}+
\pfrac{H_{\e}}{t}\dfrac{t}{s}+\pfrac{H_{\e}}{\Hv}\dfrac{\Hv}{s}\\
&=\dfrac{q^{i}}{s}\dfrac{p_{i}}{s}-\dfrac{p_{i}}{s}\dfrac{q^{i}}{s}+
\dfrac{\Hv}{s}\dfrac{t}{s}-\dfrac{t}{s}\dfrac{\Hv}{s}\\
&\equiv0.
\end{align*}
Thus, if $\Hv(0)=\Hv_{0}$ is identified with the system's initial energy
$\Hv_{0}=H(\bq_{0},\bp_{0},0)$ at $t=0$, then the condition
$H_{\e}(\bq,\bp,t,\Hv)=0$, $\d H_{\e}(\bq,\bp,t,\Hv)/\d s=0$ is
\emph{automatically} fulfilled along the system's trajectory that is given by
the solution of the extended set of canonical equations~(\ref{caneq-def}).

The extended phase-space variable $\Hv(s)$ is defined as the
particular function of the independent variable, $s$, that
represents the \emph{value} of the conventional Hamiltonian, $H$.
In accordance with Eqs.~(\ref{p0-def0}) and (\ref{hamid}),
we thus determine $H$ for any given extended Hamiltonian
$H_{\e}$ by solving $H_{\e}=0$ for $\Hv$.
Then, $H$ emerges as the right-hand side of the equation $\Hv=H$.

In the converse case, if the conventional Hamiltonian $H$ is
given and $H_{\e}$ is set up according to Eq.~(\ref{H1-def}), then
the canonical equation for $\d t/\d s$ yields an \emph{identity},
hence allows arbitrary parametrizations of time,
$$
\dfrac{t}{s}=-\pfrac{H_{\e}}{\Hv}=-\pfrac{}{\Hv}\left[\big(H(\bq,\bp,t)-
\Hv\big)\dfrac{t}{s}\right]=\dfrac{t}{s}.
$$
Exactly as in the Lagrangian description, this is not astonishing
as a conventional Hamiltonian $H$ generally does not provide the
information for an equation of motion for $t(s)$, i.e., for a
particular parametrization of time $t$.
Furthermore, setting up the extended Hamiltonian $H_{\e}$ according
to Eq.~(\ref{H1-def}) on the basis of  a given conventional Hamiltonian
$H$ does not generate additional information on the actual dynamical system.

Corresponding to Eq.~(\ref{p0-def0}), we may introduce the variable
$\Hv_{\e}$ as the \emph{value} of the extended Hamiltonian $H_{\e}$.
We can formally imagine $H_{\e}$ to be also a function of $s$ in
addition to its dependence of the extended phase-space variables,
\begin{equation}\label{p0-def1}
\Hv_{\e}\stackrel{\not\equiv}{=}H_{\e}(\bq,\bp,t,\Hv,s).
\end{equation}
By virtue of the extended set of canonical
equations~(\ref{caneq-def}), we find that $\Hv_{\e}$ is a
constant of motion if and only if $H_{\e}$ does \emph{not}
explicitly depend on $s$,
$$
\Hv_{\e}(s)=\mathrm{const.}\qquad\Longleftrightarrow\qquad
H_{\e}=H_{\e}(\bq,\bp,t,\Hv).
$$
In this case, $s$ can be regarded as a \emph{cyclic variable}, with
$\Hv_{\e}$ the pertaining constant of motion, and hence its conjugate.
Thus, in the same way as $(\Hv,t)$ constitutes a pair of
canonically conjugate variables, so does the pair $(\Hv_{\e},s)$,
i.e., the \emph{value} $\Hv_{\e}$ of the extended Hamiltonian
$H_{\e}$ and the parameterization of the system's variables
in terms of $s$.
In the context of a corresponding quantum description,
this additional pair of canonically conjugate variables
gives rise to the additional uncertainty relation
\begin{equation}\label{uncertain}
\Delta\Hv_{\e}\,\Delta s\geq\onehalf\hbar.
\end{equation}
Thus, in a quantum system whose classical limit is described
by an extended Hamiltonian $H_{\e}$, we cannot simultaneously
measure exactly both a deviation $\Delta\Hv_{\e}$ from the
hypersurface condition $\Delta\Hv_{\e}(s)=0$ from Eqs.~(\ref{hamid}),
(\ref{p0-def1}) \emph{and} the actual value of the system
evolution parameter $s$.
For the particular extended Hamiltonian $H_{\e}$ of a relativistic
particle in an external electromagnetic field, to be discussed
in Sect.~\ref{sec:ham1-em}, the condition reflects the
\emph{relativistic energy-momentum correlation}, whereas the
parameter $s$ represents the particle's \emph{proper time}.
For this particular system, the uncertainty
relation~(\ref{uncertain}) thus states the we cannot have
simultaneous knowledge on a deviation from the relativistic
energy-momentum correlation~(\ref{constraint-em})
\emph{and} the particle's proper time.
The extended Lagrangian $L_{\e}$ and the uncertainty
relation~(\ref{uncertain}) constitute together the cornerstones
for deriving the \emph{relativistic generalization} of Feynman's
path integral approach to non-relativistic quantum physics,
to be presented in Sect.~\ref{sec:pathint}.

To end this section, we remark that the extended Hamiltonian
$H_{\e}$ most frequently found in literature is given by (cf, for
instance, Refs.~\cite{lanczos,siegel,thirring,stiefel,tsiga,synge})
\begin{equation}\label{H1-triv}
H_{\e}(\bq,\bp,t,\Hv)=H(\bq,\bp,t)-\Hv.
\end{equation}
According to Eqs.~(\ref{caneq-def1}), the canonical
equation for $\d t/\d s$ is obtained as
$$
\dfrac{t}{s}=-\pfrac{H_{\e}}{\Hv}=1.
$$
Up to arbitrary shifts of the origin of our time scale,
we thus \emph{identify} $t(s)$ with $s$.
As all other partial derivatives of $H_{\e}$ coincide with
those of $H$, so do the respective canonical equations.
The system description in terms of $H_{\e}$ from Eq.~(\ref{H1-triv})
is thus \emph{identical} to the conventional description
and does not provide any additional information.
The extended Hamiltonian~(\ref{H1-triv}) thus constitutes
the simplest form of a \emph{trivial extended Hamiltonian}.
\subsection{\label{sec:cantra}Extended canonical transformations}
The conventional theory of canonical transformations is built upon
the conventional action integral from Eq.~(\ref{principle0}).
In this theory, the Newtonian absolute time $t$ plays the role
\emph{of the common independent variable} of both original
and destination system.
Similarly to the conventional theory, we may build the
\emph{extended theory of canonical equations} on the basis
of the extended action integral from Eq.~(\ref{principle1}).
With the time $t=q^{0}/c$ and the configuration space variables
$q^{i}$ treated on equal footing, we are enabled to correlate
two Hamiltonian systems, $H$ and $\HC$, with different
time scales, $t(s)$ and $T(s)$, hence to canonically map
the system's time $t$ and its conjugate quantity $\Hv$ in addition
to the mapping of generalized coordinates $\bq$ and momenta $\bp$.
The global timelike evolution parameter $s$ then plays the role of
the common independent variable of both systems, $H$ and $\HC$.
A general mapping of all dependent variables may be
formally expressed as
\begin{equation}\label{can1}
Q^{\mu}=Q^{\mu}(q^{\nu},p_{\nu}),\qquad
P_{\mu}=P_{\mu}(q^{\nu},p_{\nu}),\qquad\mu=0,\ldots,n
\end{equation}
Completely parallel to the conventional theory, the subgroup of
general transformations~(\ref{can1}) that satisfy the principle
$\delta S_{\e}=0$ of the action functional~(\ref{principle1})
is referred to as ``canonical'',
\begin{equation}\label{canbed2a}
\delta\int_{s_{a}}^{s_{b}}
L_{\e}\left(q^{\nu},\dfrac{q^{\nu}}{s}\right)\d s
=\delta\int_{s_{a}}^{s_{b}}
L_{\e}^{\prime}\left(Q^{\nu},\dfrac{Q^{\mu}}{s}\right)\d s.
\end{equation}
The action integrals may be expressed
equivalently in terms of an extended Hamiltonian by means
of the Legendre transformation~(\ref{legendre1}).
We thus get the following condition for a
transformation~(\ref{can1}) to be canonical
\begin{equation}\label{canbed2}
\delta\int_{s_{a}}^{s_{b}}\left[
\sum_{\mu=0}^{n}p_{\mu}\dfrac{q^{\mu}}{s}-
H_{\e}\big(q^{\nu},p_{\nu}\big)\right]\d s
=\delta\int_{s_{a}}^{s_{b}}\left[
\sum_{\mu=0}^{n}P_{\mu}\dfrac{Q^{\mu}}{s}-
\HC_{\e}\big(Q^{\nu},P_{\nu}\big)\right]\d s.
\end{equation}
As we are operating with \emph{functionals}, the
conditions~(\ref{canbed2a}) and (\ref{canbed2}) hold if the
\emph{integrands} differ at most by the derivative $\d\FC_{1}/\d s$
of an arbitrary differentiable function $\FC_{1}(q^{\nu},Q^{\nu})$
\begin{align}
L_{\e}&=L_{\e}^{\prime}+\dfrac{\FC_{1}}{s}\label{can2a}\\
\sum_{\mu=0}^{n}p_{\mu}\dfrac{q^{\mu}}{s}-H_{\e}&=
\sum_{\mu=0}^{n}P_{\mu}\dfrac{Q^{\mu}}{s}-\HC_{\e}+\dfrac{\FC_{1}}{s}.
\label{can2}
\end{align}
Because of
$$
\delta\int_{s_{a}}^{s_{b}}\dfrac{\FC_{1}}{s}\,\d s=
\delta\left({\left.\FC_{1}\right|}_{s_{b}}\right)-
\delta\left({\left.\FC_{1}\right|}_{s_{a}}\right)\equiv0,
$$
a term $\d\FC_{1}/\d s$ does not contribute to the variation
of the action functional~(\ref{principle1}).
This means that the particular path
$\left(\bar{\bq}(s),\bar{t}(s)\right)$ on which the
action integral takes on an extremum is maintained.

We restrict ourselves to functions $\FC_{1}(q^{\nu},Q^{\nu})$
of the old and the new extended configuration space variables,
hence to a function of those variables, whose derivatives
match those of the integrands in Eq.~(\ref{canbed2}).
Calculating the $s$-derivative of $\FC_{1}$,
\begin{equation}\label{genf1}
\dfrac{\FC_{1}}{s}=\sum_{\mu=0}^{n}\left[
\pfrac{\FC_{1}}{q^{\mu}}\dfrac{q^{\mu}}{s}+
\pfrac{\FC_{1}}{Q^{\mu}}\dfrac{Q^{\mu}}{s}\right],
\end{equation}
we then get \emph{unique} transformation rules by comparing the
coefficients of Eq.~(\ref{genf1}) with those of (\ref{can2})
\begin{equation}\label{F1}
p_{\mu}=\pfrac{\FC_{1}}{q^{\mu}},\qquad
P_{\mu}=-\pfrac{\FC_{1}}{Q^{\mu}},\qquad
\HC_{\e}=H_{\e}.
\end{equation}
$\FC_{1}$ is referred to as the \emph{extended generating function}
of the --- now generalized --- canonical transformation.
The extended Hamiltonian $H_{\e}$ has the important property
that its \emph{value} is conserved under extended canonical
transformations.
This means that the system's physical evolution is kept
being confined to the surface $\HC_{\e}=0$, hence that
the condition~$(\ref{hamid})$ is maintained in the
transformed system, as required.
Corresponding to the extended set of canonical equations,
the additional transformation rule is given for the index $\mu=0$.
This transformation rule may be expressed equivalently
in terms of $t(s)$, $\Hv(s)$, and $T(s)$, $\HCv(s)$ as
\begin{equation}\label{F1a}
\Hv=-\pfrac{\FC_{1}}{t},\qquad\HCv=\pfrac{\FC_{1}}{T},
\end{equation}
with $\HCv$, correspondingly to Eq.~(\ref{p0-def0}),
the value of the transformed Hamiltonian $\HC$
\begin{equation}\label{P0-def0}
P_{0}(s)=-\frac{\HCv(s)}{c},\qquad\HCv(s)\stackrel{\not\equiv}{=}
\HC(\bQ(s),\bP(s),T(s)).
\end{equation}
The addressed transformed Hamiltonian $\HC$ is finally obtained from the
general correlation of conventional and extended Hamiltonians from
Eq.~(\ref{H1-def}), and the transformation rule $\HC_{\e}=H_{\e}$
for the extended Hamiltonian from Eq.~(\ref{F1})
$$
\Big[\HC(\bQ,\bP,T)-\HCv\Big]\dfrac{T}{s}=
\Big[H(\bq,\bp,t)-\Hv\Big]\dfrac{t}{s}.
$$
Eliminating the evolution parameter $s$, we arrive at the
following two equivalent transformation rules for the
conventional Hamiltonians under extended canonical transformations
\begin{align}
\Big[\HC(\bQ,\bP,T)-\HCv\Big]\pfrac{T}{t}&=H(\bq,\bp,t)-\Hv\nonumber\\
\Big[H(\bq,\bp,t)-\Hv\Big]\pfrac{t}{T}&=\HC(\bQ,\bP,T)-\HCv.
\label{canham1}
\end{align}
The transformation rules (\ref{canham1}) are generalizations
of the rule for conventional canonical transformations as now
cases with $T\ne t$ are included.
We will see at the end of this section that the rules~(\ref{canham1})
merge for the particular case $T=t$ into the corresponding rules
of conventional canonical transformation theory.

By means of the Legendre transformation
\begin{equation}\label{legendre-F1}
\FC_{2}(q^{\nu},P_{\nu})=\FC_{1}(q^{\nu},Q^{\nu})+
\sum_{\mu=0}^{n}Q^{\mu}P_{\mu},\qquad P_{\mu}=-\pfrac{\FC_{1}}{Q^{\mu}},
\end{equation}
we may express the extended generating function of a generalized
canonical transformation equivalently as a function of the
original extended configuration space variables $q^{\nu}$
and the extended set of transformed canonical momenta $P_{\nu}$.
As, by definition, the functions $\FC_{1}$ and $\FC_{2}$
agree in their dependence on the $q^{\mu}$, so do the
corresponding transformation rules
$$
\pfrac{\FC_{1}}{q^{\mu}}=\pfrac{\FC_{2}}{q^{\mu}}=p_{\mu}.
$$
This means that all $q^{\mu}$ do not take part in the transformation
defined by~(\ref{legendre-F1}).
As $\FC_{1}$ does not depend on the $P_{\nu}$, the new transformation
rule pertaining to $\FC_{2}$ thus follows immediately as
\begin{align*}
\pfrac{\FC_{2}}{P_{\nu}}&=\sum_{\mu=0}^{n}Q^{\mu}
\pfrac{P_{\mu}}{P_{\nu}}=\sum_{\mu=0}^{n}Q^{\mu}\delta_{\mu}^{\nu}\\
&=Q^{\nu}.
\end{align*}
The new set of transformation rules, which is, of course, equivalent
to the previous set from Eq.~(\ref{F1}), is thus
\begin{equation}\label{F2}
p_{\mu}=\pfrac{\FC_{2}}{q^{\mu}},\qquad
Q^{\mu}=\pfrac{\FC_{2}}{P_{\mu}},\qquad
\HC_{\e}=H_{\e}.
\end{equation}
Expressed in terms of the variables $\bq$, $\bp$, $t$, $\Hv$, and
$\bQ$, $\bP$, $T$, $\HCv$
the new set of coordinate transformation rules takes
on the more elaborate form
\begin{equation}\label{rules}
p_{i}=\pfrac{\FC_{2}}{q^{i}},\qquad Q^{i}=\pfrac{\FC_{2}}{P_{i}},\qquad
\Hv=-\pfrac{\FC_{2}}{t},\qquad T=-\pfrac{\FC_{2}}{\HCv}.
\end{equation}
Similarly to the conventional theory of canonical transformations,
there are two more possibilities to define a generating function
of an extended canonical transformation.
By means of the Legendre transformation
$$
\FC_{3}(p_{\nu},Q^{\nu})=\FC_{1}(q^{\nu},Q^{\nu})-
\sum_{\mu=0}^{n}q^{\mu}p_{\mu},\qquad
p_{\mu}=-\pfrac{\FC_{1}}{q^{\mu}},
$$
we find in the same manner as above the transformation rules
\begin{equation}\label{F3}
q^{\mu}=-\pfrac{\FC_{3}}{p_{\mu}},\qquad
P_{\mu}=-\pfrac{\FC_{3}}{Q^{\mu}},\qquad
\HC_{\e}=H_{\e}.
\end{equation}
Finally, applying the Legendre transformation, defined by
$$
\FC_{4}(p_{\nu},P_{\nu})=\FC_{3}(p_{\nu},Q^{\nu})+
\sum_{\mu=0}^{n}Q^{\mu}P_{\mu},\qquad
P_{\mu}=-\pfrac{\FC_{3}}{Q^{\mu}},
$$
the following equivalent version of transformation rules emerges
$$
q^{\mu}=-\pfrac{\FC_{4}}{p_{\mu}},\qquad
Q^{\mu}= \pfrac{\FC_{4}}{P_{\mu}},\qquad
\HC_{\e}=H_{\e}.
$$
Calculating the second derivatives of the generating functions,
we conclude that the following correlations for the derivatives
of the general mapping from Eq.~(\ref{can1}) must hold
for the entire set of extended phase-space variables,
$$
\pfrac{Q^{\mu}}{q^{\nu}}=\pfrac{p_{\nu}}{P_{\mu}},\qquad
\pfrac{Q^{\mu}}{p_{\nu}}=-\pfrac{q^{\nu}}{P_{\mu}},\qquad
\pfrac{P_{\mu}}{q^{\nu}}=-\pfrac{p_{\nu}}{Q^{\mu}},\qquad
\pfrac{P_{\mu}}{p_{\nu}}=\pfrac{q^{\nu}}{Q^{\mu}}.
$$
Exactly if these conditions are fulfilled for all
$\mu,\nu=0,\ldots,n$, then the extended coordinate
transformation~(\ref{can1}) is canonical and preserves the
form of the extended set of canonical equations~(\ref{caneq-def}).
Otherwise, we are dealing with a general, non-canonical coordinate
transformation that does \emph{not} preserve the form of the
canonical equations.

The connection of the extended canonical transformation theory
with the conventional one is furnished by the particular extended
generating function
\begin{equation}\label{F2-triv}
\FC_{2}(\bq,\bP,t,\HCv)=f_{2}(\bq,\bP,t)-t\HCv,
\end{equation}
with $f_{2}(\bq,\bP,t)$ denoting a conventional generating function.
According to Eqs.~(\ref{rules}), the coordinate transformation
rules following from~(\ref{F2-triv}) are
$$
p_{i}=\pfrac{f_{2}}{q^{i}},\qquad Q^{i}=\pfrac{f_{2}}{P_{i}},
\qquad\Hv=-\pfrac{f_{2}}{t}+\HCv,\qquad T=t.
$$
With $\partial T/\partial t=1$, the general transformation
rule~(\ref{canham1}) for conventional Hamiltonians now yields
the well-known rule for Hamiltonians $\HC$ under conventional
canonical transformations,
$$
\HC(\bQ,\bP,t)=H(\bq,\bp,t)+\HCv-\Hv=H(\bq,\bp,t)+\pfrac{f_{2}}{t}.
$$
Canonical transformations that are defined by extended generating
functions of the form of Eq.~(\ref{F2-triv}) leave the time variable
unchanged and thus define the subgroup of conventional canonical
transformations within the general group of extended canonical transformations.
Corresponding to the trivial extended Hamiltonian from
Eq.~(\ref{H1-triv}), we may refer to (\ref{F2-triv}) as
the \emph{trivial extended generating function}.
\subsection{\label{sec:hj}Extended Hamilton-Jacobi equation}
In the context of the extended canonical transformation theory,
we may derive an extended version of the Hamilton-Jacobi equation.
We are looking for a generating function $\FC_{2}(q^{\nu},P_{\nu})$
of an extended canonical transformation that maps a given extended
Hamiltonian $H_{\e}=0$ into a transformed extended Hamiltonian
$\HC_{\e}=0$ with the property that \emph{all} partial
derivatives of $\HC_{\e}(Q^{\nu},P_{\nu})$ vanish.
Hence, according to the extended set of canonical equations~(\ref{caneq-def}),
the derivatives of all canonical variables $Q^{\mu}(s),P_{\mu}(s)$
with respect to the system's evolution parameter $s$ must vanish
\begin{equation}\label{hj-cond}
\pfrac{\HC_{\e}}{P_{\mu}}=\dfrac{Q^{\mu}}{s}=0,\qquad
-\pfrac{\HC_{\e}}{Q^{\mu}}=\dfrac{P_{\mu}}{s}=0,\qquad
\mu=0,\ldots,n.
\end{equation}
This means that \emph{all} transformed canonical
variables $Q^{\mu},P_{\mu}$ must be constants of motion.
Writing the variables for the index $\mu=0$ separately, we thus have
$$
T=\mathrm{const.},\quad Q^{i}=\mathrm{const.},\quad
\HCv=\mathrm{const.},\quad P_{i}=\mathrm{const.}
$$
Thus, corresponding to the conventional Hamilton-Jacobi formalism,
the vectors of the transformed canonical variables, $\bQ$ and $\bP$,
are constant.
Yet, in the extended formalism, the transformed time $T$
is also a constant.
The particular generating function $\FC_{2}(q^{\nu},P_{\nu})$ that
defines transformation rules for the extended set of canonical variables
such that Eqs.~(\ref{hj-cond}) hold for the transformed variables
thus defines a mapping of the entire system into its state at a
fixed instant of time, hence --- up to trivial shifts in the origin
of the time scale --- into its initial state at $T=t(0)$
$$
T=t(0),\quad Q^{i}=q^{i}(0),\quad P_{i}=p_{i}(0),\quad
\HCv=H(\bq(0),\bp(0),t(0)).
$$
We may refer to this particular generating function
as the \emph{extended Hamiltonian action function}
$\FC_{2}\equiv S_{\e}(q^{\nu},P_{\nu})$.
According to the transformation rule $\HC_{\e}=H_{\e}$ for extended
Hamiltonians from Eq.~(\ref{F1}), we obtain the transformed
extended Hamiltonian $\HC_{\e}\equiv0$ simply by expressing the
original extended Hamiltonian $H_{\e}=0$ in terms of the
transformed variables.
This means for the conventional Hamiltonian $H(\bq,\bp,t)$
according to Eq.~(\ref{H1-def}) in conjunction with
the transformation rules from Eqs.~(\ref{rules}),
$$
\left[H\left(\bq,\pfrac{S_{\e}}{\bq},t\right)+
\pfrac{S_{\e}}{t}\right]\dfrac{t}{s}=0.
$$
As we have $\d s/\d t\ne0$ in general, we finally get
the generalized form of the Hamilton-Jacobi equation,
\begin{equation}\label{gen-hjgl}
H\left(q^{1},\ldots,q^{n},\pfrac{S_{\e}}{q^{1}},\ldots,
\pfrac{S_{\e}}{q^{n}},t\right)+\pfrac{S_{\e}}{t}=0.
\end{equation}
Equation (\ref{gen-hjgl}) has exactly the \emph{form} of
the conventional Hamilton-Jacobi equation.
Yet, it is actually a \emph{generalization} as the extended
action function $S_{\e}$ represents an \emph{extended}
generating function of type $\FC_{2}$, as defined by
Eq.~(\ref{legendre-F1}).
This means that $S_{\e}$ is also a function of the
(constant) transformed energy $\HCv=-P(0)$.

Summarizing, the extended Hamilton-Jacobi equation may be
interpreted as defining the mapping of all canonical
coordinates $\bq$, $\bp$, $t$, and $\Hv$ of the actual
system into constants $\bQ$, $\bP$, $T$, and $\HCv$.
In other words, it defines the mapping of the entire dynamical
system from its actual state at time $t$ into its state at a
\emph{fixed instant of time}, $T$, which could be the initial
conditions.
\subsection{\label{sec:pathint}Generalized path integral
with extended Lagrangians}
In Feynman's path integral approach to quantum mechanics,
the space and time evolution of a wave function $\psi(\bq,t)$
is formulated in terms of a transition amplitude density
$K(b,a)$, also referred to as a \emph{kernel}, or, a \emph{propagator}:
\begin{equation}\label{wave-evol}
\psi(\bq_{b},t_{b})=\int_{-\infty}^{\infty}
K(\bq_{b},t_{b};\bq_{a},t_{a})\,\psi(\bq_{a},t_{a})\,\d^3\bq_{a}.
\end{equation}
The parameterized kernel $K_{\sigma}(b,a)$ for a parameterized action $S_{\e}$
is given by the multiple path integral
\begin{equation}\label{kernel-para}
K_{\sigma}(b,a)=\iint\exp\left\{\frac{i}{\hbar}
S_{\e}[\bq(s),t(s)]\right\}\mathscr{D}^3\bq(s)\mathscr{D}t(s).
\end{equation}
Herein, the integrals are to be taken over all paths that go
from $(\bq_{a},t_{a})$ at $s_{a}$ to $(\bq_{b},t_{b})$ at $s_{b}$.
The justification for integrating over all times is that in
relativistic physics we must treat space and time on equal footing.
Hence, we must allow the laboratory time $t$ to take any value ---
negative and even positive ones --- if we regard $t$ from the
viewpoint of a particle with its proper time $s$.
We thus additionally integrate over all \emph{histories} of the particle.
The integration over all \emph{futures} can then be interpreted as
integration over all histories of the \emph{anti-particle}, whose
proper timescale runs backwards in terms of the particle's proper timescale.\cite{stueckel}

If the time paths and the spatial paths are taken to be
independent of each other, hence if we do not incorporate
the shell condition (\ref{lagid}) into the integration boundaries,
we also sum over all particles off the mass shell.
The action functional $S_{\e}$ stands for the $s$-integral over the
extended Lagrangian $L_{\e}$, as defined by Eq.~(\ref{principle1}).

In classical dynamics, the parameterization of space and time
variables can be eliminated by means of the shell condition~(\ref{lagid}).
For the corresponding quantum description, the uncertainty
principle from Eq.~(\ref{uncertain}) applies.
It tells us that an \emph{accurate} fulfillment of the
condition $\Delta\Hv_{\e}(s)=0$ is related to a
\emph{complete uncertainty} about the parameterization
of the system's variables phase-space in terms of $s$.
Therefore, in the context of the path integral approach, the condition
$\Delta\Hv_{\e}(s)=0$ is incorporated by \emph{integrating} the parameterized
kernel $K_{\sigma}(b,a)$ over all possible parameterizations
$\sigma=s_{b}-s_{a}>0$ of coordinates $\bq(s)$ and time~$t(s)$.
The final kernel, hence the transition amplitude density is thus given by
\begin{equation}\label{kernel-gen}
K(b,a)=\int_{0}^{\infty}K_{\sigma}(b,a)\,\d\sigma.
\end{equation}
This means that \emph{all} parameterized kernels
$K_{\sigma}(b,a)$ contribute with \emph{equal weight}
to the total transition amplitude $K(b,a)$.
As an example, we calculate in Sect.~\ref{sec:prop} the
explicit form of the space-time propagator for the wave
function of a relativistic free particle from the extended
Lagrangian $L_{\e}$ of the pertaining classical system.

For an infinitesimal step $\delta\epsilon=s_{b}-s_{a}$, we may approximate
the action functional $S_{\e}$ from Eq.~(\ref{principle1}) by
$$
S_{\e,\delta\epsilon}[q^{\mu}(s)]=\delta\epsilon\,L_{\e}\left(\frac{q^{\mu}_{b}+
q^{\mu}_{a}}{2},\frac{q^{\mu}_{b}-q^{\mu}_{a}}{\delta\epsilon}\right).
$$
For $s_{b}=s_{a}+\delta\epsilon$, the kernel
$K_{\sigma}(s_{a}+\delta\epsilon,s_{a})$
from Eq.~(\ref{kernel-para}) that yields the transition amplitude
density for a particle along this infinitesimal interval
$s_{b}-s_{a}$ is accordingly given by
$$
K(b,a)=\frac{1}{M}
\exp\left[\frac{i}{\hbar}S_{\e,\delta\epsilon}\right].
$$
As we proceed an infinitesimal step $\delta\epsilon$ only, and
then take the limit $\delta\epsilon\to0$, the integration~(\ref{kernel-gen})
over all possible parameterizations of this step must be omitted.
For, conversely to the situation discussed beforehand, a small
$\delta\epsilon=\Delta s$ is related to a large uncertainty with
respect to satisfying the condition $\Delta\Hv_{\e}(s)=0$, so that in the
limit $\delta\epsilon\to0$ the condition ceases to exist.

The yet to be determined normalization factor
$M$ represents the integration measure for one step
of the multiple path integral~(\ref{kernel-para}).
Clearly, this measure must depend on the step size~$\delta\epsilon$.
The transition of a given wave function $\psi(q^{\mu}_{a})$
at the particle's proper time $s_{a}$ to the wave function
$\psi(q^{\mu}_{b})$ that is separated by an infinitesimal
proper time interval $\delta\epsilon=s_{b}-s_{a}$ can now
be formulated according to Eq.~(\ref{wave-evol}) as
\begin{equation}\label{trans-infini}
\psi(q^{\mu}_{b})=\frac{1}{M}\int\exp\left[
\frac{i}{\hbar}S_{\e,\delta\epsilon}\right]\psi(q^{\mu}_{a})\,\d^{4}q_{a}.
\end{equation}
Note that we integrate here over the entire space-time.
To serve as test for this approach, we derive
in Sect.~\ref{sec:kg} the Klein-Gordon equation on the basis of
the extended Lagrangian $L_{\e}$ for a relativistic particle
in an external electromagnetic field.
\section{Examples of extended Hamilton-Lagrange systems}
\subsection{\label{sec:lag1-fp}Extended Lagrangian
for a relativistic free particle}
As only expressions of the form $\bq^{2}-c^{2}t^{2}$ are preserved
under the Lorentz group, the conventional Lagrangian
for a \emph{free point particle} of mass $m$, given by
\begin{equation}\label{lagnr-fp}
L^{\mathrm{nr}}\left(\bq,\dfrac{\bq}{t},t\right)=T-V=
\onehalf m{\left(\dfrac{\bq}{t}\right)}^{2}-mc^{2},
\end{equation}
is obviously not Lorentz-invariant.
Yet, in the extended description, a corresponding Lorentz-invariant
Lagrangian $L_{\e}$ can be constructed by introducing $s$ as the new
independent variable, and by treating the space and time variables,
$\bq(s)$ and $q^{0}=ct(s)$ equally.
This is achieved by adding the corresponding derivative of the
time variable $t(s)$,
\begin{equation}\label{lag1-fp}
L_{\e}\left(\bq,\dfrac{\bq}{s},t,\dfrac{t}{s}\right)=
\onehalf mc^{2}\left[{\frac{1}{c^{2}}\left(\dfrac{\bq}{s}\right)}^{2}-
{\left(\dfrac{t}{s}\right)}^{2}-1\right].
\end{equation}
The constant third term has been defined accordingly to ensure that
$L_{\e}$ converges to $L^{\mathrm{nr}}$ in the limit $\d t/\d s\to1$.
Of course, the dynamics following from (\ref{lagnr-fp}) and
(\ref{lag1-fp}) are \emph{different} --- which reflects the
modification our dynamics encounters if we switch
from a non-relativistic to a relativistic description.
The Lagrangian~(\ref{lag1-fp}) is no homogeneous form of first
order in the velocities $\d q^{\mu}/\d s,\mu=0,\ldots,3$.
Therefore, we obtain from Eq.~(\ref{lagid}) the hypersurface condition, also referred to as the mass shell condition:
\begin{equation}\label{constraint-lag}
\frac{1}{c^{2}}{\left(\dfrac{\bq}{s}\right)}^{2}-{\left(\dfrac{t}{s}\right)}^{2}+1=0\qquad\Leftrightarrow\qquad
\frac{1}{c^{2}}{\left(\dfrac{\bq}{t}\right)}^{2}+{\left(\dfrac{s}{t}\right)}^{2}-1=0.
\end{equation}
We thus encounter the reciprocal value of the relativistic scale factor, $\gamma$,
\begin{equation}\label{gamma-inv}
\dfrac{s}{t}=\sqrt{1-\frac{1}{c^{2}}{\left(\dfrac{\bq}{t}\right)}^{2}}=\gamma^{-1},
\end{equation}
which shows that in the case of the Lagrangian~(\ref{lag1-fp}) the system evolution
parameter $s$ is physically nothing else than the particle's proper time.
Inserting the condition~(\ref{constraint-lag}) into the Lagrangian yields the constant \emph{value} of $L_{\e}$,
$$
L_{\e}\big|_{\text{mass shell}}=-mc^{2}.
$$
In contrast to the non-relativistic description,
the constant rest energy term $-\onehalf mc^{2}$ in the
extended Lagrangian~(\ref{lag1-fp}) is essential.
Consequently, the extended Lagrangian~(\ref{lag1-fp}) is no homogeneous
form of first order in the velocities $\d q^{\mu}/\d s,\mu=0,\ldots,3$,
the condition~(\ref{lagid}) is not satisfied \emph{identically}.
Yet, in the derivation of~(\ref{lagid}), we have assumed that
a corresponding conventional Lagrangian $L$ exists, hence
a Lagrangian that depends on $\d t/\d s$ only indirectly
via the reparameterization condition
$$
\dfrac{\bq}{t}=\dfrac{\bq/\d s}{t/\d s}
$$
from Eq.~(\ref{lag1}) applied to its velocities.
We must, therefore, make sure that such a corresponding
\emph{conventional} Lagrangian $L$ exists, hence a function
$L=L_{\e}\,\d s/\d t$ that does not depend anymore on $s$.
For the extended Lagrangian $L_{\e}$ from Eq.~(\ref{lag1-fp}),
a corresponding conventional Lagrangian $L$ indeed exists.
Inserting Eq.~(\ref{constraint-lag}) into Eq.~(\ref{lag1-fp}),
we find with Eq.~(\ref{gamma-inv})
\begin{align}
L\left(\bq,\dfrac{\bq}{t},t\right)&=\left.
L_{\e}\left(\bq,\dfrac{\bq}{s},t,\dfrac{t}{s}\right)\right|_{\text{mass shell}}\dfrac{s}{t}\nonumber\\
&=-mc^{2}\,\dfrac{s}{t}\nonumber\\
&=-mc^{2}\sqrt{1-\frac{1}{c^{2}}{\left(
\dfrac{\bq}{t}\right)}^{2}}.
\label{lag-fp}
\end{align}
We thus encounter the well-known conventional Lagrangian
of a relativistic free particle.
In contrast to the equivalent extended Lagrangian from
Eq.~(\ref{lag1-fp}), the Lagrangian~(\ref{lag-fp}) is \emph{not}
quadratic in the derivatives of the dependent variables, $\bq(t)$.
The loss of the quadratic form originates from
the \emph{projection} of the hypersurface description within the
tangent bundle $T(\MB\times\RB)$ to the description within $(T\MB)\times\RB$.
The quadratic form is recovered in the non-relativistic limit
by expanding the square root, which yields the Lagrangian
$L^{\mathrm{nr}}$ from Eq.~(\ref{lagnr-fp}).

Denoting by $q^{\mu}$ the components of the contravariant four-vector
of space-time variables $(q^{0},\ldots,q^{3})=(ct,x,y,z)$,
the corresponding covariant vector is then
$(q_{0},\ldots,q_{3})=(-ct,x,y,z)$ for the metric
$\eta_{\mu\nu}=\mathrm{diag}(-1,1,1,1)$.
Adopting the ``summation convention,'' which means to sum over
all quantities with pairs of identical covariant and contravariant
indices, the non-homogeneous extended Lagrangian from
Eq.~(\ref{lag1-fp}) can then be rewritten in covariant notation as
$$
L_{\e}\left(q^{\mu},\dfrac{q^{\mu}}{s}\right)=
\onehalf m\left(\dfrac{q^{\alpha}}{s}\dfrac{q_{\alpha}}{s}-c^{2}\right).
$$
The hypersurface condition~(\ref{lagid}) is then expressed as
$$
\dfrac{q^{\alpha}}{s}\dfrac{q_{\alpha}}{s}=-c^{2},
$$
which depicts the constant length of the four-velocity vector.

To summarize, with $L_{\e}$ from Eq.~(\ref{lag1-fp}), we have found
a \emph{non-trivial} extended Lagrangian $L_{\e}$, i.e.\ an extended
Lagrangian that is \emph{non-homogeneous} in its velocities
and possesses a corresponding conventional Lagrangian
$L=L_{\e}\,\d s/\d t$, with $\d s/\d t$ determined by Eq.~(\ref{lagid})
that now embodies an implicit equation rather than an identity.
In addition to the equations of motion for $\bq(s)$, this $L_{\e}$
determines uniquely the correlation $t(s)$ of the laboratory time
$t$ to the particle's proper time,~$s$.
\subsection{\label{sec:lag1-triv-fp}Trivial extended Lagrangian
for a relativistic free particle}
Given the conventional Lagrangian~(\ref{lag-fp}), we may
immediately set up the corresponding \emph{trivial} extended
Lagrangian according to Eq.~(\ref{lag1}) by multiplying
$L$ with $\d t/\d s$
\begin{align}
L_{\e}^{\mathrm{triv}}\left(\bq,\dfrac{\bq}{s},t,\dfrac{t}{s}\right)&=
-mc\sqrt{c^{2}{\left(\dfrac{t}{s}\right)}^{2}-
{\left(\dfrac{\bq}{s}\right)}^{2}}\nonumber\\
&=-mc\sqrt{-\dfrac{q^{\alpha}}{s}\dfrac{q_{\alpha}}{s}}.
\label{eq:lag1-triv-fp}
\end{align}
We easily convince ourselves that the trivial extended Lagrangian
satisfies Eq.~(\ref{lagid}) \emph{identically}
\begin{align*}
\pfrac{L_{\e}^{\mathrm{triv}}}{\left(\dfrac{q^{\mu}}{s}\right)}\dfrac{q^{\mu}}{s}&=
\frac{mc}{\sqrt{-\dfrac{q^{\alpha}}{s}\dfrac{q_{\alpha}}{s}}}\dfrac{q_{\mu}}{s}\dfrac{q^{\mu}}{s}\\
&=-mc\sqrt{-\dfrac{q^{\alpha}}{s}\dfrac{q_{\alpha}}{s}}\\
&\equiv L_{\e}^{\mathrm{triv}}
\end{align*}
and thus fulfills Eq.~(\ref{lagid-deri}),
$$
\pfrac{^{2}L_{\e}^{\mathrm{triv}}}
{\left(\dfrac{q^{\mu}}{s}\right)\partial\left(\dfrac{q_{\nu}}{s}\right)}=
\frac{mc}{{\left(-\dfrac{q^{\alpha}}{s}\dfrac{q_{\alpha}}{s}\right)}^{3/2}}
\left(\dfrac{q^{\nu}}{s}\dfrac{q_{\mu}}{s}-\delta_{\mu}^{\nu}\,
\dfrac{q^{\beta}}{s}\dfrac{q_{\beta}}{s}\right),
$$
hence
\begin{align*}
\pfrac{^{2}L_{\e}^{\mathrm{triv}}}
{\left(\dfrac{q^{\mu}}{s}\right)\partial\left(\dfrac{q_{\nu}}{s}\right)}\dfrac{q^{\mu}}{s}&=
mc{{\left(-\dfrac{q^{\alpha}}{s}\dfrac{q_{\alpha}}{s}\right)}^{-\frac{3}{2}}}
\left(\dfrac{q^{\nu}}{s}\dfrac{q_{\mu}}{s}-\delta_{\mu}^{\nu}\,
\dfrac{q^{\beta}}{s}\dfrac{q_{\beta}}{s}\right)\dfrac{q^{\mu}}{s}\\
&\equiv0.
\end{align*}
The subsequent equation of motion for $t(s)$ does \emph{not}
determine a parametrization of time $t$ but rather allows for
arbitrary parametrizations.
As a trivial extended Lagrangian $L_{\e}^{\mathrm{triv}}$ generally
follows by multiplying a given conventional Lagrangian $L$ by $\d t/\d s$,
a formally covariant description is encountered in the sense that space
and time variables are then treated on equal footing.
Yet, no additional information on the dynamical system is
provided by the transition from $L$ to $L_{\e}^{\mathrm{triv}}$.
\subsection{\label{sec:ham1-triv-fp}Trivial extended Hamiltonian
for a relativistic free particle}
For a trivial extended Lagrangian, is not possible to derive the
corresponding trivial extended Hamiltonian as the Legendre
transformation of a homogeneous extended Lagrangian is singular.
This does not mean that a corresponding extended Hamiltonian does
not exist, as it is frequently claimed in literature\cite{johns}.
To the contrary, for any conventional Lagrangian $L$ that can be
Legendre-transformed into a corresponding conventional Hamiltonian $H$,
one can always set up $L_{\e}$ according to Eq.~(\ref{lag1}) and $H_{\e}$
according to Eq.~(\ref{H1-def}).
Setting up the extended set of Euler-Lagrange equations for
a trivial extended Lagrangian then yields exactly the same
description of the given dynamical system as setting up the
extended set of canonical equations for the trivial extended
Hamiltonian obtained this way.

In order to set up the trivial extended Hamiltonian $H_{\e}^{\mathrm{triv}}$
that corresponds to the trivial extended Lagrangian $L_{\e}^{\mathrm{triv}}$
from Eq.~(\ref{eq:lag1-triv-fp}) of the free relativistic point particle,
one must first Legendre-transform the underlying conventional
Lagrangian~(\ref{lag-fp}) to the corresponding conventional Hamiltonian
according to
$$
H(\bq,\bp,t)=\bp\dot{\bq}-L(\bq,\dot{\bq},t),\qquad\bp=\pfrac{L}{\dot{\bq}}.
$$
For the particular Lagrangian~(\ref{lag-fp}), one finds
$$
\bp=\frac{m\dot{\bq}}{\sqrt{1-\frac{\dot{\bq}^{2}}{c^{2}}}}\quad\Rightarrow\quad
H=\frac{mc^{2}}{\sqrt{1-\frac{\dot{\bq}^{2}}{c^{2}}}}.
$$
A Hamiltonian must be expressed in terms of the canonical momenta
rather than by the velocities, hence $\dot{\bq}$ must be expressed in terms of $\bp$,
$$
H^{2}=\frac{m^{2}c^{4}}{1-\frac{\dot{\bq}^{2}}{c^{2}}},\qquad
\bp^{2}=\frac{m^{2}\dot{\bq}^{2}}{1-\frac{\dot{\bq}^{2}}{c^{2}}}\quad\Rightarrow\quad
H^{2}-\bp^{2}c^{2}=\frac{m^{2}c^{4}}{1-\frac{\dot{\bq}^{2}}{c^{2}}}
\left(1-\frac{\dot{\bq}^{2}}{c^{2}}\right)=m^{2}c^{4},
$$
hence
$$
H(\bq,\bp,t)=\sqrt{\bp^{2}c^{2}+m^{2}c^{4}}.
$$
The corresponding trivial extended Hamiltonian $H_{\e}^{\mathrm{triv}}$
can now be set up according to the general recipe from Eq.~(\ref{H1-def})
\begin{equation}\label{eq:ham1-triv-fp}
H_{\e}^{\mathrm{triv}}(\bq,\bp,t,\Hv)=\left(\sqrt{\bp^{2}c^{2}+m^{2}c^{4}}-\Hv\right)\dfrac{t}{s}.
\end{equation}
In contrast to the Lagrangian description, the factor $\d t/\d s$ does
not represent a conjugate variable but enters into the canonical
equations as an external factor.
The trivial extended Hamiltonian~(\ref{eq:ham1-triv-fp}) has exactly
the same information content on the underlying dynamical system as
the trivial extended Lagrangian from~(\ref{eq:lag1-triv-fp}) and
thus yields identical equations of motion.
In particular, $H_{\e}^{\mathrm{triv}}$ equally does \emph{not}
determine a parametrization of time, $t=t(s)$, but rather allows for
arbitrary parametrizations.
This can be seen by setting up the respective canonical equation
$$
\dfrac{t}{s}=-\pfrac{H_{\e}^{\mathrm{triv}}}{\Hv}=\dfrac{t}{s}.
$$
One thus finds an \emph{identity} but no substantial canonical equation for $t=t(s)$.
\subsection{\label{sec:lag1-em}Extended Lagrangian for a
relativistic particle in an external electromagnetic field}
The non-homogeneous extended Lagrangian $L_{\e}$ of a point particle of
mass $m$ and charge $\zeta$ in an external electromagnetic field
that is described by the potentials $(\phi,\bA)$ is given by
\begin{align}
L_{\e}\!\left(\bq,\dfrac{\bq}{s},t,\dfrac{t}{s}\right)=
\onehalf mc^{2}\!\left[{\frac{1}{c^{2}}\left(\dfrac{\bq}{s}\right)}^{2}\!-
{\left(\dfrac{t}{s}\right)}^{2}\!-1\right]\!+\frac{\zeta}{c}\bA(\bq,t)
\dfrac{\bq}{s}-\zeta\,\phi(\bq,t)\dfrac{t}{s}.\nonumber\\
\label{lag1-em}
\end{align}
The associated hypersurface condition~(\ref{lagid}) for $L_{\e}$ coincides with that
for the free-particle Lagrangian from Eq.~(\ref{constraint-lag})
as all terms linear in the velocities drop out
\begin{equation}\label{hypersurface-em}
{\left(\dfrac{t}{s}\right)}^{2}-\frac{1}{c^{2}}
{\left(\dfrac{\bq}{s}\right)}^{2}-1=0.
\end{equation}
Similar to the free particle case from Eq.~(\ref{lag-fp}), the extended
Lagrangian~(\ref{lag1-em}) may be projected into $(T\MB)\times\RB$
to yield the well-known conventional relativistic Lagrangian $L$
\begin{equation}\label{lagr-em}
L\left(\bq,\dfrac{\bq}{t},t\right)=
-mc^{2}\sqrt{1-\frac{1}{c^{2}}{\left(
\dfrac{\bq}{t}\right)}^{2}}+\frac{\zeta}{c}\bA
\dfrac{\bq}{t}-\zeta\,\phi.
\end{equation}
Again, the quadratic form of the velocity terms
is lost owing to the projection.

For small velocity $\d\bq/\d t$, the quadratic form is regained
as the square root in (\ref{lagr-em}) may be expanded to yield
the conventional non-relativistic Lagrangian for a point particle
in an external electromagnetic field,
\begin{equation}\label{lag-em}
L^{\mathrm{nr}}\left(\bq,\dfrac{\bq}{t},t\right)=
\onehalf m{\left(\dfrac{\bq}{t}\right)}^{2}+\frac{\zeta}{c}\bA
\dfrac{\bq}{t}-\zeta\,\phi-mc^{2}.
\end{equation}
Significantly, this Lagrangian can be derived \emph{directly},
hence without the detour over the projected
Lagrangian~(\ref{lagr-em}), from the extended
Lagrangian~(\ref{lag1-em}) by letting $\d t/\d s\to1$.

It is instructive to review the Lagrangian~(\ref{lag1-em})
and its non-relativistic limit~(\ref{lag-em})
in covariant notation.
With Einstein's summation convention and the notation
$A_{0}(q^{\mu})=-\phi(q^{\mu})$ for the particular constant metric
$\eta_{\mu\nu}=\mathrm{diag}(-1,1,1,1)$,
the extended Lagrangian~(\ref{lag1-em}) then writes
\begin{equation}\label{lag1-em2}
L_{\e}\left(q^{\mu},\dfrac{q^{\mu}}{s}\right)=
\onehalf m\,\eta_{\alpha\beta}\dfrac{q^{\alpha}}{s}\dfrac{q^{\beta}}{s}+
\frac{\zeta}{c}A_{\alpha}\dfrac{q^{\alpha}}{s}-\onehalf mc^{2}.
\end{equation}
The hypersurface condition~(\ref{hypersurface-em}) is then converted into
\begin{equation}\label{hypersurface-em2}
\eta_{\alpha\beta}\dfrac{q^{\alpha}}{s}\dfrac{q^{\beta}}{s}=-c^{2}.
\end{equation}
Correspondingly, the non-relativistic Lagrangian~(\ref{lag-em})
has the equivalent representation
\begin{equation}\label{lag-em2}
L^{\mathrm{nr}}\left(q^{\mu},\dfrac{q^{\mu}}{t}\right)=
\onehalf m\,\eta_{\alpha\beta}\dfrac{q^{\alpha}}{t}\dfrac{q^{\beta}}{t}+
\frac{\zeta}{c}A_{\alpha}\dfrac{q^{\alpha}}{t}-\onehalf mc^{2}.
\end{equation}
Note that $(\d q^{0}/\d t)(\d q_{0}/\d t)=-c^{2}$, which yields
the second half of the rest energy term, so that (\ref{lag-em2})
indeed agrees with (\ref{lag-em}).
Comparing the Lagrangian~(\ref{lag-em2}) with the extended Lagrangian
from Eq.~(\ref{lag1-em2}) --- and correspondingly the Lagrangians
(\ref{lag1-em}) and (\ref{lag-em}) --- we notice that the
transition to the non-relativistic description is made by
identifying the proper time $s$ with the laboratory time $t=q^{0}/c$.
The remarkable formal similarity of the Lorentz-invariant
extended Lagrangian~(\ref{lag1-em2}) with the non-invariant
conventional Lagrangian~(\ref{lag-em2}) suggests that approaches based
on non-relativistic Lagrangians $L^{\mathrm{nr}}$ may be transposed to
a relativistic description by (i) introducing the proper time $s$ as
the new system evolution parameter, (ii) treating the time $t(s)$
as an \emph{additional dependent variable} on equal footing with
the configuration space variables $\bq(s)$ --- commonly referred to
as the ``principle of homogeneity in space-time'' --- and (iii) by
replacing the conventional non-relativistic Lagrangian $L^{\mathrm{nr}}$
with the corresponding Lorentz-invariant extended Lagrangian $L_{\e}$,
similar to the transition from (\ref{lag-em2}) to (\ref{lag1-em2}).
\subsection{\label{sec:ham1-em}Extended Hamiltonian for
a relativistic particle in an external electromagnetic field}
The extended Hamiltonian counterpart $H_{\e}$ of the non-homogeneous extended
Lagrangian~(\ref{lag1-em}) for a relativistic point particle
in an external electromagnetic field is obtained via the
Legendre transformation prescription from Eqs.~(\ref{legendre1})
and (\ref{p-def}).
The transition to the extended Hamiltonian $H_{\e}$
is easiest calculated by starting form the covariant
form~(\ref{lag1-em2}) of $L_{\e}$ and afterwards converting
the results to $3$-vector notation.
According to Eqs.~(\ref{p-def}) and (\ref{p0-def}),
the canonical momenta $p_{\mu}$ are introduced by
\begin{equation}\label{p-em}
p_{\mu}=\pfrac{L_{\e}}{\left(\dfrac{q^{\mu}}{s}\right)}=
m\,\eta_{\mu\alpha}\dfrac{q^{\alpha}}{s}+\frac{\zeta}{c}A_{\mu}=
p_{\mu,\rk}+\frac{\zeta}{c}A_{\mu}.
\end{equation}
We notice that the \emph{kinetic} momentum
$p_{\mu,\rk}=m\,\d q_{\mu}/\d s$ differs
from the \emph{canonical} momentum $p_{\mu}$
in the case of a non-vanishing external potential $A_{\mu}\neq0$.
The condition for the Legendre transform of $L_{\e}$
to exist is that its Hesse matrix $\partial^{2}L_{\e}/%
[\partial(\d q^{\mu}/\d s)\partial(\d q_{\nu}/d s)]$
must be non-singular, hence that the determinant of this
matrix does not vanish.
For the Lagrangian $L_{\e}$ from Eq.~(\ref{lag1-em2}),
this is actually the case as
$$
\det\left(\pfrac{^{2}L_{\e}}{\left(\dfrac{q^{\mu}}{s}\right)
\partial\left(\dfrac{q_{\nu}}{s}\right)}\right)=
\det\left(m\,\delta_{\mu}^{\nu}\right)=
m^{4}\ne0.
$$
This falsifies claims made in literature\cite{johns} that the
Hesse matrix associated with an extended Lagrangian $L_{\e}$
be \emph{generally singular}, and that for this reason an extended
Hamiltonian $H_{\e}$ \emph{generally} could not be obtained
by a Legendre transformation of an extended Lagrangian $L_{\e}$.
The necessary condition for an extended Hamiltonian $H_{\e}$ to emerge
form a Legendre transformation of an extended Lagrangian $L_{\e}$
is that $L_{\e}$ must not be a \emph{homogeneous} function
of first order in its velocities $\d q^{\mu}/\d s$.

With the Hesse condition being actually satisfied,
the extended Hamiltonian $H_{\e}$ that follows as the
Legendre transform~(\ref{legendre1}) of $L_{\e}$ reads
\begin{align*}
\qquad H_{\e}(q^{\mu},p_{\mu})&=\dfrac{q^{\alpha}}{s}\left(
m\dfrac{q_{\alpha}}{s}+\frac{\zeta}{c}A_{\alpha}\right)-
\onehalf m\dfrac{q^{\alpha}}{s}\dfrac{q_{\alpha}}{s}-
\frac{\zeta}{c}A_{\alpha}\dfrac{q^{\alpha}}{s}+\onehalf mc^{2}\\
&=\onehalf m\dfrac{q^{\alpha}}{s}\dfrac{q_{\alpha}}{s}+\onehalf mc^{2}.
\end{align*}
As any Hamiltonian must be expressed in terms of the canonical
momenta rather than through velocities, $H_{\e}$ takes on the
more elaborate final form according to Eq.~(\ref{p-em})
\begin{equation}\label{h1-em}
H_{\e}(q^{\mu},p_{\mu})=\frac{1}{2m}
\left(p_{\alpha}-\frac{\zeta}{c}A_{\alpha}\right)
\left(p^{\alpha}-\frac{\zeta}{c}A^{\alpha}\right)+\onehalf mc^{2}.
\end{equation}
This extended Hamiltonian coincides with the ``super-Hamiltonian''
that was postulated by Misner, Thorne, and Wheeler\cite{misner}.

In covariant notation, the condition $H_{\e}=0$ thus follows as
$$
\left(p_{\alpha}-\frac{\zeta}{c}A_{\alpha}\right)
\left(p^{\alpha}-\frac{\zeta}{c}A^{\alpha}\right)+m^{2}c^{2}=0,
$$
which follows equivalently if the velocities in the hypersurface
condition~(\ref{hypersurface-em2}) are replaced by the canonical
momenta according to Eq.~(\ref{p-em}).
In terms of the conventional $3$-vectors for the canonical momentum
$\bp$ and vector potential $\bA$, and the scalars, energy $\Hv$ and
electric potential $\phi$, the extended Hamiltonian $H_{\e}$ is
equivalently expressed as
\begin{equation}\label{h1-em2}
\qquad H_{\e}(\bq,\bp,t,\Hv)=\frac{1}{2m}
\left[{\left(\bp-\frac{\zeta}{c}\bA(\bq,t)\right)}^{2}-
{\left(\frac{\Hv-\zeta\phi(\bq,t)}{c}\right)}^{2}\right]+
\onehalf mc^{2},
\end{equation}
and the condition $H_{\e}=0$ furnishes the usual
relativistic energy relation
\begin{equation}\label{constraint-em}
{\big(\Hv-\zeta\phi(\bq,t)\big)}^{2}=c^{2}
{\left(\bp-\frac{\zeta}{c}\bA(\bq,t)\right)}^{2}+m^{2}c^{4}.
\end{equation}
The \emph{conventional} Hamiltonian $H$
that describes the same dynamics is determined according
to Eq.~(\ref{p0-def0}) as the particular \emph{function},
whose \emph{value} coincides with $\Hv$.
Solving $H_{\e}=0$ from Eq.~(\ref{h1-em2}) for $\Hv$, we
directly find $H$ as the left-hand side of the equation $H=\Hv$,
\begin{equation}\label{h-em}
H(\bq,\bp,t)=\sqrt{c^{2}{\left(\bp-\frac{\zeta}{c}\bA(\bq,t)\right)}^{2}+
m^{2}c^{4}}+\zeta\phi(\bq,t)=\Hv.
\end{equation}
The conventional Hamiltonian $H^{\mathrm{nr}}$ that describes the
particle dynamics in the non-relativistic limit is obtained
from the Lorentz-invariant Hamiltonian~(\ref{h-em}) by
expanding the square root
$$
H^{\mathrm{nr}}(\bq,\bp,t)=\frac{1}{2m}{\left(\bp-\frac{\zeta}{c}\bA(\bq,t)
\right)}^{2}+\zeta\phi(\bq,t)+mc^{2}.
$$
In contrast to the extended Lagrangian description, a \emph{direct}
way to transpose the relativistic extended Hamiltonian from
Eq.~(\ref{h1-em2}) into the non-relativistic Hamiltonian
$H^{\mathrm{nr}}$ does not exist.
We conclude that the Lagrangian approach is more appropriate
if we want to ``translate'' a given non-relativistic
Hamilton-Lagrange system into the corresponding
Lorentz-invariant description.

In order to show that the extended Hamiltonian~(\ref{h1-em2}) and the
well-known conventional Hamiltonian~(\ref{h-em}) indeed yield the same dynamics,
we now set up the extended set of canonical equations~(\ref{caneq-def})
for the covariant extended Hamiltonian~(\ref{h1-em})
\begin{align}
-\pfrac{H_{\e}}{q^{\mu}}=\dfrac{p_{\mu}}{s}&=
\frac{\zeta}{mc}\eta^{\alpha\beta}\left(p_{\alpha}-\frac{\zeta}{c}A_{\alpha}\right)
\pfrac{A_{\alpha}}{q^{\mu}}\nonumber\\
\hphantom{-}\pfrac{H_{\e}}{p_{\mu}}=\dfrac{q^{\mu}}{s}&=
\frac{1}{m}\eta^{\mu\alpha}\left(p_{\alpha}-\frac{\zeta}{c}A_{\alpha}\right).\label{eqmo-em}
\end{align}
In the notation of scalars and $3$-vectors, the pair of
equations~(\ref{eqmo-em}) separates into the following
equivalent set of four equations
\begin{align}
\dfrac{p_{i}}{s}&=
\frac{\zeta}{mc}\left(p_j-\frac{\zeta}{c}A_j\right)
\pfrac{A^j}{q^{i}}-\frac{\zeta}{mc^{2}}
\left(\Hv-\zeta\phi\right)\pfrac{\phi}{q^{i}}\nonumber\\
\dfrac{\Hv}{s}&=
-\frac{\zeta}{mc}\left(p_j-\frac{\zeta}{c}A_j\right)
\pfrac{A^j}{t}+\frac{\zeta}{mc^{2}}
\left(\Hv-\zeta\phi\right)\pfrac{\phi}{t}\nonumber\\
\dfrac{q^{i}}{s}&=
\frac{1}{m}\left(p^{i}-\frac{\zeta}{c}A^{i}\right)\nonumber\\
\dfrac{t}{s}&=
\frac{1}{mc^{2}}\left(\Hv-\zeta\phi\right).
\label{eqmo-em2}
\end{align}
From the last equation, we deduce the derivative of the
inverse function $s=s(t)$ and insert the condition from
Eq.~(\ref{constraint-em})
\begin{equation}\label{ds-dt}
\dfrac{s}{t}=\frac{mc^{2}}{\Hv-\zeta\phi}=
\frac{mc^{2}}{\sqrt{c^{2}{\left(\bp-\frac{\zeta}{c}
\bA(\bq,t)\right)}^{2}+m^{2}c^{4}}}.
\end{equation}
The canonical equations~(\ref{eqmo-em2}) can now be expressed
equivalently with the time $t$ as the independent variable
\begin{align}
-\dfrac{p_{i}}{t}&=-\dfrac{p_{i}}{s}\dfrac{s}{t}=
-\frac{\zeta c}{\sqrt{c^{2}{\left(\bp-\frac{\zeta}{c}
\bA(\bq,t)\right)}^{2}+m^{2}c^{4}}}
\left(\bp-\frac{\zeta}{c}\bA\right)
\pfrac{\bA}{q^{i}}+\zeta\pfrac{\phi}{q^{i}}\nonumber\\
\hphantom{-}\dfrac{\Hv}{t}&=\hphantom{-}\dfrac{\Hv}{s}\dfrac{s}{t}=
-\frac{\zeta c}{\sqrt{c^{2}{\left(\bp-\frac{\zeta}{c}
\bA(\bq,t)\right)}^{2}+m^{2}c^{4}}}\left(\bp-\frac{\zeta}{c}\bA\right)
\pfrac{\bA}{t}+\zeta\pfrac{\phi}{t}\nonumber\\
\hphantom{-}\dfrac{q^{i}}{t}&=\hphantom{-}\dfrac{q^{i}}{s}\dfrac{s}{t}=
\frac{c^{2}}{\sqrt{c^{2}{\left(\bp-\frac{\zeta}{c}
\bA(\bq,t)\right)}^{2}+m^{2}c^{4}}}
\left(p^{i}-\frac{\zeta}{c}A^{i}\right).
\label{eqmo-em3}
\end{align}
The right-hand sides of Eqs.~(\ref{eqmo-em3}) are
exactly the partial derivatives $\partial H/\partial q^{i}$,
$\partial H/\partial t$, and $\partial H/\partial p_{i}$
of the Hamiltonian~(\ref{h-em}) --- and hence its canonical
equations, which was to be shown.

The physical meaning of the $\d t/\d s$ is worked out by
casting it to the equivalent form
$$
\dfrac{t}{s}=\sqrt{1+\frac{{\left(\bp-\frac{\zeta}{c}
\bA(\bq,t)\right)}^{2}}{m^{2}c^{2}}}=
\sqrt{1+{\left(\frac{\bp_{\rk}(s)}{mc}\right)}^{2}}=\gamma(s),
$$
with $\bp_{\rk}(s)$ the instantaneous \emph{kinetic}
momentum of the particle.
The dimensionless quantity $\d t/\d s$ thus represents the
instantaneous value of the relativistic scale factor $\gamma$.
\subsection{\label{sec:lortra}Lorentz transformation
as an extended canonical transformation}
We know that the Lorentz transformation provides the
rules according to which a physical system is transformed
from one inertial reference system into an other.
On the other hand, a mapping of one Hamiltonian into another
is constituted by a canonical transformation.
Consequently, the Lorentz transformation must be a
particular canonical transformation.
As the Lorentz transformation \emph{always} involves
a transformation of the time scales \mbox{$t\mapsto T$},
this transformation can only be represented by an
\emph{extended} canonical transformation.
Its generating function $\FC_{2}$ is given by
\begin{align}
\FC_{2}(\bq,\bP_{\rk},t,\HCv_{\rk})=\bP_{\rk}\bq-
\gamma\!\left[\HCv_{\rk}t+\bbeta\!\left(
\bP_{\rk}ct-\frac{\HCv_{\rk}}{c}\,\bq\right)\!\right]\!+
\frac{\gamma-1}{\beta^{2}}\big(\bbeta\bP_{\rk}\big)
\big(\bbeta\bq\big)\nonumber\\
\label{gen-lorentz}
\end{align}
with $\bbeta=\bv/c$ the constant vector that delineates
the scaled relative velocity $\bv$ of both reference systems,
and $\gamma$ the dimensionless relativistic
scale factor $\gamma=1/\sqrt{1-\bbeta^{2}}$.
In order to also cover cases where the particle moves within an
external potential, the index ``kin'' indicates that the momenta
and the energy are to be understood as the ``kinetic'' quantities,
as defined in Eq.~(\ref{p-em}).
The generating function~(\ref{gen-lorentz}) generalizes the
free-particle generator presented earlier in Ref.~\cite{struck}.
The general transformation rules~(\ref{rules}) for extended
generating functions of type $\FC_{2}$ yield for the particular
generator from Eq.~(\ref{gen-lorentz})
\begin{align*}
\bp_{\rk}=\pfrac{\FC_{2}}{\bq}&=\bP_{\rk}+
\frac{\gamma\bbeta}{c}\,\HCv_{\rk}+\frac{\gamma-1}{\beta^{2}}
\bbeta\big(\bbeta\bP_{\rk}\big),\,
\Hv_{\rk}=-\pfrac{\FC_{2}}{t}=\gamma
\HCv_{\rk}+c\gamma\bbeta\bP_{\rk},\\
\bQ=\pfrac{\FC_{2}}{\bP_{\rk}}&=\bq-
\gamma\bbeta\,ct+\frac{\gamma-1}{\beta^{2}}
\bbeta\big(\bbeta\bq\big),\qquad\qquad\;\;\:
\,T=-\pfrac{\FC_{2}}{\HCv_{\rk}}=\gamma t-
\frac{\gamma}{c}\,\bbeta\bq.
\end{align*}
In matrix form, the transformation rules for the space-time
coordinates, $\bQ$ and $T$, are
\begin{equation}\label{lorentz-rules1}
\begin{pmatrix}
\bQ\\ cT
\end{pmatrix}
=
\begin{pmatrix}
1+\left(\frac{\gamma-1}{\beta^{2}}
\bbeta\right)\bbeta & \quad & -\gamma\bbeta\;\\
-\gamma\bbeta & \quad & \gamma
\end{pmatrix}
\begin{pmatrix}
\bq\\ ct
\end{pmatrix}.
\end{equation}
The corresponding linear relation for the kinetic momentum
vector $\bp_{\rk}$ and the kinetic energy $\Hv_{\rk}$ is
\begin{equation}\label{lorentz-rules2}
\begin{pmatrix}
\bp_{\rk}\\ \Hv_{\rk}/c
\end{pmatrix}
=
\begin{pmatrix}
1+\left(\frac{\gamma-1}{\beta^{2}}
\bbeta\right)\bbeta & \quad & \gamma\bbeta\;\\
\gamma\bbeta & \quad & \gamma
\end{pmatrix}
\begin{pmatrix}
\bP_{\rk}\\ \HCv_{\rk}/c
\end{pmatrix}.
\end{equation}
If we replace the kinetic momenta with the canonical momenta
according to Eq.~\!(\ref{p-em}), it is not astonishing to find
that the external potentials obey the same transformation rule
as the momenta,
$$
\begin{pmatrix}
\bA\\ \phi
\end{pmatrix}
=
\begin{pmatrix}
1+\left(\frac{\gamma-1}{\beta^{2}}
\bbeta\right)\bbeta & \quad & \gamma\bbeta\;\\
\gamma\bbeta & \quad & \gamma
\end{pmatrix}
\begin{pmatrix}
\bA^{\prime}\\ \phi^{\prime}
\end{pmatrix}.
$$
We easily convince ourselves that the transformation~(\ref{lorentz-rules1})
preserves the condition (\ref{constraint-lag}) that equally
applies for a particle in an external potential.
Correspondingly, the transformation (\ref{lorentz-rules2})
preserves the conditions (\ref{constraint-em}).
As a consequence, we have established  the important result
that the extended Hamiltonian $H_{\e}$ from Eq.~(\ref{h1-em2}) is
also preserved under Lorentz transformations
$$
\HC_{\e}(\bP,\bQ,T,\HCv)=H_{\e}(\bp,\bq,t,\Hv).
$$
This is in agreement with the general canonical transformation
rule for extended Hamiltonians from Eq.~(\ref{F1})

According to the subsequent rule for the conventional
Hamiltonians, $H$ and $\HC$, from Eq.~(\ref{canham1}),
and $\partial T/\partial t=\gamma$, we find
\begin{equation}\label{canham2}
\big(\HC-\HCv_{\rk}\big)\gamma=H-\Hv_{\rk}.
\end{equation}
In conjunction with the energy transformation rule
from Eq.~(\ref{lorentz-rules2}),
$\Hv_{\rk}=\gamma\HCv_{\rk}+\bbeta\gamma\bP_{\rk} c$,
we get from Eq.~(\ref{canham2}) the transformation rule
for a Hamiltonian $H$ under Lorentz transformations
$$
H=\gamma\big(\HC+\bbeta c\bP_{\rk}\big).
$$
As expected, the Hamiltonians, $H$ and $\HC$, transform
equally as their respective values, $\Hv_{\rk}$ and $\HCv_{\rk}$.
\subsection{\label{ex:gen-noether}
Infinitesimal canonical transformations,
generalized Noether theorem}
A general infinitesimal extended transformation is generated by
\begin{equation}\label{gen-infini}
\FC_{2}(q^{\nu},P_{\nu})=\sum_{\alpha=0}^{n}q^{\alpha}P_{\alpha}+
\delta\epsilon\,I(q^{\nu},p_{\nu}).
\end{equation}
In this generating function, $\delta\epsilon\in\RB$ denotes an
infinitesimal parameter, whereas the differentiable function
$I(q^{\nu},p_{\nu})$ quantifies the deviation of the actual
infinitesimal transformation from the \emph{identity}.
We first derive the coordinate transformation
rules for the particular generating function~(\ref{gen-infini})
according to the general rules~(\ref{F2}),
\begin{align}
p_{\mu}&=\pfrac{\FC_{2}}{q^{\mu}}=P_{\mu}+
\delta\epsilon\,\pfrac{I}{q^{\mu}},\nonumber\\
Q^{\mu}&=\pfrac{\FC_{2}}{P_{\mu}}=q^{\mu}\,+
\delta\epsilon\,\pfrac{I}{P_{\mu}},\label{genrules-infini}\\
H_{\e}^{\prime}&=H_{\e}\nonumber.
\end{align}
To first order in $\delta\epsilon$, the variations $\delta p_{\mu}$,
$\delta q^{\mu}$, and $\delta H_{\e}$ are obtained from the
transformation rules~(\ref{genrules-infini}) as
\begin{eqnarray}
\delta p_{\mu}&\equiv&
P_{\mu}-p_{\mu}\,\,=-\delta\epsilon\,\pfrac{I}{q^{\mu}},\nonumber\\
\delta q^{\mu}&\equiv&
Q^{\mu}-q^{\mu}\;=\hphantom{-}\delta\epsilon\,\pfrac{I}{p_{\mu}},
\label{rules-infini}\\
\delta H_{\e}&\equiv&H_{\e}^{\prime}-H_{\e}=0.\nonumber
\end{eqnarray}
Obviously, any function $I(q^{\nu},p_{\nu})$ is \emph{invariant}
under the infinitesimal transformation it defines,
$$
\delta I=\sum_{\alpha=0}^{n}\left(
\pfrac{I}{q^{\alpha}}\,\delta q^{\alpha}+
\pfrac{I}{p_{\alpha}}\,\delta p_{\alpha}\right)=
\delta\epsilon\sum_{\alpha=0}^{n}\left(
\pfrac{I}{q^{\alpha}}\,\pfrac{I}{p_{\alpha}}-
\pfrac{I}{p_{\alpha}}\,\pfrac{I}{q^{\alpha}}\right)\equiv0.
$$
This is \emph{not} necessarily true for the extended Hamiltonian $H_{\e}$.
The condition $H_{\e}=0$ from Eq.~(\ref{hamid}) enters into the extended
canonical transformation theory in the way that we must \emph{explicitly verify}
that $H_{\e}^{\prime}=H_{\e}$ actually holds under the transformation rules
of the canonical variables that are defined by the generating function.
Only then the physical motion of the transformed system keeps being
confined to the phase-space surface $H_{\e}^{\prime}=0$, as required
for the system to be \emph{physical}.
In the case of the \emph{infinitesimal}
transformation~(\ref{rules-infini}), the transformation rule for the
extended Hamiltonian $H_{\e}$ is satisfied exactly if $\delta H_{\e}=0$
under the infinitesimal variations of the canonical variables.
For the transformation rules~(\ref{rules-infini}), the variation
of $H_{\e}$ due to the variations $\delta q^{\nu}$ and
$\delta p_{\nu }$ of the canonical variables is given by
\begin{align*}
\delta H_{\e}&=\sum_{\alpha=0}^{n}\left(
\pfrac{H_{\e}}{q^{\alpha}}\,\delta q^{\alpha}+
\pfrac{H_{\e}}{p_{\alpha}}\,\delta p_{\alpha}\right)\\
&=\delta\epsilon\sum_{\alpha=0}^{n}\left(
\pfrac{H_{\e}}{q^{\alpha}}\pfrac{I}{p_{\alpha}}-
\pfrac{H_{\e}}{p_{\alpha}}\pfrac{I}{q^{\alpha}}\right)\\
&=\delta\epsilon\,{\left[H_{\e},I\right]}_{\text{ext}},
\end{align*}
with the last expression defining the extended Poisson bracket.
Thus, the canonical transformation rule $\delta H_{\e}=0$
from Eqs.~(\ref{rules-infini}) is actually fulfilled
if and only if the characteristic function
$I(q^{\nu},p_{\nu})$ in~(\ref{gen-infini}) satisfies
\begin{equation}\label{noetherinvariant}
\sum_{\alpha=0}^{n}\left(\pfrac{I}{q^{\alpha}}\pfrac{H_{\e}}{p_{\alpha}}-
\pfrac{I}{p_{\alpha}}\pfrac{H_{\e}}{q^{\alpha}}\right)=
{\left[I,H_{\e}\right]}_{\text{ext}}=0.
\end{equation}
Along the system trajectory, the canonical equations~(\ref{caneq-def}) apply.
As a consequence, the partial derivatives of $H_{\e}$
in~(\ref{noetherinvariant}) may be replaced accordingly to yield
\begin{equation}\label{invar-g}
\sum_{\alpha=0}^{n}\left(\pfrac{I}{q^{\alpha}}\dfrac{q^{\alpha}}{s}+
\pfrac{I}{p_{\alpha}}\dfrac{p_{\alpha}}{s}\right)=\dfrac{I}{s}=0.
\end{equation}
Thus, $I(q^{\nu},p_{\nu})$ must ``commute'' with the extended
Hamiltonian $H_{\e}$, hence must be \emph{invariant} along
the system's phase-space trajectory in order for the
transformation~(\ref{gen-infini}) to comply with the requirement
$\delta H_{\e}=0$ for an extended canonical transformation.
Then and only then the generating function~(\ref{gen-infini})
defines an extended \emph{canonical} transformation and thus
ensures the action functional~(\ref{canbed2a}) to be preserved.
The correlation~(\ref{invar-g}) of a system invariant $I$
to a transformation that preserves the action functional
--- hence to a \emph{canonical} transformation --- establishes
the most general form of Noether's theorem in the realm of the
extended Hamilton-Lagrange formulation of point mechanics,
\begin{equation}\label{gen-noether}
{\left[I,H_{\e}\right]}_{\text{ext}}=0\quad\Longleftrightarrow\quad
\dfrac{I}{s}=0\quad\Longleftrightarrow\quad\delta H_{\e}=0.
\end{equation}
We may rewrite the condition~(\ref{noetherinvariant}) in terms
of a conventional Hamiltonian $H$ if we distinguish the space
coordinates $q^{i},\;i=1,\ldots,n$ from the time coordinate $t$.
With the replacements $q^{0}=ct, p_{0}=-\Hv/c$, $\Hv$ denoting the
instantaneous \emph{value} of the conventional Hamiltonian $H$, and
$$
\pfrac{H_{\e}}{t}=\pfrac{H}{t}\dfrac{t}{s},\qquad
\pfrac{H_{\e}}{\Hv}=-\dfrac{t}{s},\qquad
\pfrac{H_{\e}}{q^{i}}=\pfrac{H}{q^{i}}\dfrac{t}{s},\qquad
\pfrac{H_{\e}}{p_{i}}=\pfrac{H}{p_{i}}\dfrac{t}{s},
$$
according to the correlation~(\ref{H1-def}) of extended and
conventional Hamiltonians, we find for $I=I(\bp,\bq,t,\Hv)$
\begin{equation}\label{noetherinvariant3}
\pfrac{I}{t}+\pfrac{I}{\Hv}\pfrac{H}{t}+
\sum_{i=1}^{n}\left(\pfrac{I}{q^{i}}\pfrac{H}{p_{i}}-
\pfrac{I}{p_{i}}\pfrac{H}{q^{i}}\right)=0.
\end{equation}
Due to the conventional canonical equations
$$
\pfrac{H}{t}=\dfrac{\Hv}{t},\qquad
\pfrac{H}{p_{i}}=\dfrac{q^{i}}{t},\qquad
\pfrac{H}{q^{i}}=-\dfrac{p_{i}}{t},
$$
Eq.~(\ref{noetherinvariant3}) is thus equivalent to
\begin{equation}\label{noetherinvariant4}
\dfrac{I}{t}=0.
\end{equation}
In this notation, the symmetry transformation
rules~(\ref{rules-infini}) pertaining to the
invariant~(\ref{noetherinvariant4}) assume the equivalent form
\begin{equation}\label{rules-infini2}
\delta p_{i}=-\delta\epsilon\,\pfrac{I}{q^{i}},\qquad
\delta q^{i}=\delta\epsilon\,\pfrac{I}{p_{i}},\qquad
\delta\Hv=\delta\epsilon\,\pfrac{I}{t},\qquad
\delta t=-\delta\epsilon\,\pfrac{I}{\Hv}.
\end{equation}
We can always eliminate or induce an $\Hv$-dependence of $I$
by inserting the conventional Hamiltonian according to $\Hv=H$.
A representation $I=I(\bp,\bq,t)$ of the invariant $I$ does
\emph{not} depend on $\Hv$, which means that $\delta t=0$.
Then, the resulting symmetry transformation does not
involve a transformation of time.
In contrast, if $I=I(\bp,\bq,t,\Hv)$, then the invariant
defines a symmetry transformation that includes a transformation
of time, $\delta t\not=0$.
Equivalent representations $I=I(\bp,\bq,t,\Hv)$ and
$I=I(\bp,\bq,t)$ of the invariant $I$ reflect the same underlying
system symmetry, yet depicted at different instants of time $t$.

Summarizing, the set of extended canonical transformations
covers \emph{all} transformations that leave the action functional
in the generalized form of Eq.~(\ref{canbed2}) invariant.
As each canonical transformation can be defined in terms of
an infinitesimal generating function $\FC_{2}$ from
Eq.~(\ref{gen-infini}), the characteristic function
$I(\bp,\bq,t,\Hv)$ that is contained in $\FC_{2}$ then
constitutes the corresponding constant of motion.
Conversely, \emph{each} invariant $I$ of a dynamical system can be
inserted into the generating function $\FC_{2}$ of the
infinitesimal canonical transformation.
The subsequent canonical transformation rules then define
the corresponding infinitesimal symmetry transformation
of the respective dynamical system.
With the extended canonical transformation approach,
we thus encounter a \emph{generalization} of Noether's
theorem in the realm of Hamiltonian point dynamics.
\subsubsection{\label{ex:Ham-Symm}
Example: Symmetry generated by the extended Hamiltonian $H_{\e}$}
A trivial yet important example of an invariant $I$ is furnished by
the extended Hamiltonian $H_{\e}$ itself
$$
\delta H_{\e}=\delta\epsilon\,{\left[H_{\e},H_{\e}\right]}_{\text{ext}}=0,
\qquad\dfrac{H_{\e}}{s}=0.
$$
The infinitesimal transformation rules~(\ref{rules-infini})
thus define a \emph{canonical} transformation.
With $\delta\epsilon=\delta s$, their explicit form is
$$
\delta p_{\mu}=-\delta\epsilon\,\pfrac{H_{\e}}{q^{\mu}}=
\dfrac{p_{\mu}}{s}\,\delta s,\qquad
\delta q^{\mu}=\delta\epsilon\,\pfrac{H_{\e}}{p_{\mu}}=
\dfrac{q^{\mu}}{s}\,\delta s.
$$
This is obviously the infinitesimal transformation that shifts the
extended set of canonical coordinates one step $\delta s$ along the
system's extended phase-space trajectory, which always resides on
the surface $H_{\e}(q^{\nu},p_{\nu})\stackrel{\not\equiv}{=}0$.
Thus, the symmetry transformation corresponding to the constant
value of $H_{\e}$ is that the system's symplectic structure
is maintained along its evolution parameter, $s$.
\subsubsection{\label{ex:tdho}
Example: Symmetry of the time-dependent harmonic
oscillator at $\delta t=0$}
The time-dependent harmonic oscillator is a simple
one-degree-of-freedom example of a non-autonomous
dynamical system, i.e., a system whose Hamiltonian
depends explicitly on the independent variable, $t$,
\begin{equation}\label{tdham}
H(q,p,t)=\onehalf p^{2}+\onehalf\omega^{2}(t)\,q^{2}.
\end{equation}
Herein, $\omega(t)$ denotes the system's time-dependent
circular frequency.
The value $\Hv$ of the Hamiltonian $H$ is thus
not a conserved quantity.
The canonical equations and the equation of motion
immediately follow as
$$
\dot{q}=\pfrac{H}{p}=p,\quad\dot{p}=-\pfrac{H}{q}=-\omega^{2}(t)\,q,
\qquad\ddot{q}+\omega^{2}(t)\,q=0.
$$
A conserved quantity $I$ for this system is constituted
by the quadratic form
\begin{equation}\label{tdinv}
I=\beta_{e}(t)\,p^{2}+2\alpha_{e}(t)\,pq+\gamma_{e}(t)\,q^{2},
\end{equation}
provided that the time functions $\beta_{e}(t)$, $\alpha_{e}(t)$, and
$\gamma_{e}(t)$ satisfy the equations
\begin{equation}\label{tdinvcond}
\onehalf\beta_{e}\ddot{\beta}_{e}-\quarter{\dot{\beta}_{e}}^{2}+
\omega^{2}(t)\,\beta_{e}^{2}=1,\qquad\dot{\beta}_{e}=-2\alpha_{e},
\qquad\beta_{e}\gamma_{e}-\alpha_{e}^{2}=1.
\end{equation}
We easily prove the invariance of $I$ directly by calculating
its total time derivative and inserting the canonical equations
and the conditions~(\ref{tdinvcond}).

Geometrically, the quadratic form~(\ref{tdinv}) represents
an ellipse centered at the origin of the $(q,p)$-phase space
with the actual coordinates $q,p$ defining its boundary,
which varies its shape but retains its area $\pi I$.
Thus, the invariant $I$ represents the conserved area of
an ellipse with time-dependent parameters $\beta_{e}(t)$,
$\alpha_{e}(t)$, and $\gamma_{e}(t)$ that passes through
$(q(t),p(t))$.

The symmetry transformation corresponding to the
invariant~(\ref{tdinv}) follows from Eqs.~(\ref{rules-infini2})
$$
\delta p=-\delta\epsilon\pfrac{I}{q}=
\delta\sigma\left(\gamma_{e}q+\alpha_{e}p\right),\quad
\delta q=\delta\epsilon\pfrac{I}{p}=
-\delta\sigma\left(\alpha_{e}q+\beta_{e}p\right),\quad\!
\delta t=-\delta\epsilon\pfrac{I}{\Hv}\!=\!0,
$$
introducing the abbreviation $-2\delta\epsilon\equiv\delta\sigma$.
In matrix notation, this infinitesimal canonical
transformation of coordinate $q$ and momentum $p$ reads
\begin{equation}\label{tdho-infini}
\begin{pmatrix}Q\\ P\end{pmatrix}=
\left[\Eins+\AB_{\delta\sigma}\right]
\begin{pmatrix}q\\ p\end{pmatrix},\qquad
\AB_{\delta\sigma}=\delta\sigma\begin{pmatrix}-\alpha_{e}&-\beta_{e}\\
\hphantom{-}\gamma_{e}&\hphantom{-}\alpha_{e}\end{pmatrix},
\end{equation}
with $\Eins$ denoting the $2\times 2$ unit matrix.
As the coefficients of $\AB_{\delta\sigma}$ do not depend on the
canonical variables $q,p$, we may directly set up the
pertaining \emph{finite} transformation.
Equation~(\ref{tdho-infini}) may be regarded as a Taylor expansion
that could by truncated after the linear term because of very small
$\delta\sigma$.
The finite transformation for arbitrary $\sigma\in\RB$
is then given by the exponential of $\AB_{\sigma}$, hence
$$
\begin{pmatrix}
Q\\ P
\end{pmatrix}
=\MB
\begin{pmatrix}
\,q\\ \,p
\end{pmatrix},
\qquad\MB=\exp{(\AB_{\sigma})}.
$$
The general scheme for deriving the matrix exponential
$\exp{(\AB)}$ for a $2\times 2$ matrix $\AB=(a_{ij}),\;i,j=1,2$
is expressed in terms of the expression $D$,
$$
D=\sqrt{\quarter{\left(a_{11}-a_{22}\right)}^{2}+a_{12}\,a_{21}}
$$
as
\begin{align}
\MB=\exp{\left(\onehalf(a_{11}+a_{22})\right)}
\begin{pmatrix}
\cosh D+\onehalf(a_{11}-a_{22})D^{-1}
\sinh D&a_{12}D^{-1}\sinh D\\[\medskipamount]
\hspace*{-22mm}a_{21}D^{-1}\sinh D&\hspace*{-22mm}
\cosh D-\onehalf(a_{11}-a_{22})D^{-1}\sinh D
\end{pmatrix}.\nonumber\\
\label{general-M}
\end{align}
For the particular matrix $\AB_{\sigma}$ from
Eq.~(\ref{tdho-infini}), we find $a_{11}+a_{22}=0$ and $D=i\sigma$.
Due to the purely imaginary $D$, the hyperbolic sine and cosine
functions in matrix exponential are thus converted into
trigonometric sines and cosines, which finally yields
\begin{equation}\label{tdho-fini}
\begin{pmatrix}
Q\\[\medskipamount]P
\end{pmatrix}
=
\begin{pmatrix}
\cos\sigma-\alpha_{e}\sin\sigma&
-\beta_{e}\sin\sigma\\[\medskipamount]
\gamma_{e}\sin\sigma&\cos\sigma+
\alpha_{e}\sin\sigma
\end{pmatrix}
\begin{pmatrix}
q\\[\medskipamount]p
\end{pmatrix}.
\end{equation}
Note that $(Q,P)$ and $(q,p)$ as well as the ellipse parameters
$\alpha_{e}$, $\beta_{e}$, and $\gamma_{e}$ refer to the same
instant of time as the actual symmetry transformation is
associated with $\delta t=0$.
The inverse transformation is then obtained as
$$
\begin{pmatrix}q\\[\medskipamount]p
\end{pmatrix}
=
\begin{pmatrix}\cos\sigma+\alpha_{e}\sin\sigma&
\beta_{e}\sin\sigma\\[\medskipamount]
-\gamma_{e}\sin\sigma&\cos\sigma-
\alpha_{e}\sin\sigma
\end{pmatrix}
\begin{pmatrix}
Q\\[\medskipamount]P
\end{pmatrix}.
$$
Inserting $q$ and $p$ as functions of $Q$ and $P$ into the
invariant~(\ref{tdinv}), we find that the representation
of $I$ retains its form in the transformed variables
$$
I=\beta_{e}(t)\,P^{2}+2\alpha_{e}(t)\,PQ+\gamma_{e}(t)\,Q^{2}.
$$
Thus, $(Q,P)$ and $(q,p)$ both lie on the same ellipse, but
shifted with respect to each other on the ellipse's perimeter.
The geometric meaning of the one-parameter symmetry transformation
$\MB$ from Eq.~(\ref{tdho-fini}) that is associated with the
invariant $I$ from Eq.~(\ref{tdinv}) is thus to map any point
on this ellipse into another point on the \emph{same} ellipse.
The free parameter $\sigma$ of the transformation group
then specifies the particular destination point $(Q,P)$
with respect to the source point, $(q,p)$.
This can be seen from the parametric representation
of the ellipse~(\ref{tdinv})
\begin{equation}\label{ellipara}
q=\sqrt{\frac{I}{\gamma_{e}}}\left(\cos\phi-\alpha_{e}\sin\phi
\right),\qquad p=\sqrt{I\gamma_{e}}\sin\phi.
\end{equation}
Letting $\phi$ run along the interval $0\leq\phi\leq 2\pi$, we perform
one turn on the ellipse's perimeter.
The symmetry transformation~(\ref{tdho-fini}) then acts on
$(q,p)$ according to
\begin{align*}
\begin{pmatrix}
Q\\[\medskipamount]P
\end{pmatrix}
&=
\begin{pmatrix}
\cos\sigma-\alpha_{e}\sin\sigma&-\beta_{e}\sin\sigma\\[\medskipamount]
\gamma_{e}\sin\sigma&\cos\sigma+\alpha_{e}\sin\sigma
\end{pmatrix}
\begin{pmatrix}
\sqrt{I/\gamma_{e}}\left(\cos\phi-\alpha_{e}\sin\phi\right)\\[\medskipamount]
\sqrt{I\gamma_{e}}\sin\phi
\end{pmatrix}\\
&=
\begin{pmatrix}
\sqrt{I/\gamma_{e}}\left(\cos(\phi+\sigma)-
\alpha_{e}\sin(\phi+\sigma)\right)\\[\medskipamount]
\sqrt{I\gamma_{e}}\sin(\phi+\sigma)
\end{pmatrix}.
\end{align*}
Thus, $(Q,P)$ is shifted counterclockwise with respect to $(q,p)$ on the
ellipse's perimeter exactly by the phase angle $\sigma$
in the parameter representation~(\ref{ellipara}).
This accounts for $\sigma$ being referred to as a ``phase advance''.
\begin{figure}
\begin{center}
\ifpdf
\includegraphics[height=.4\linewidth]{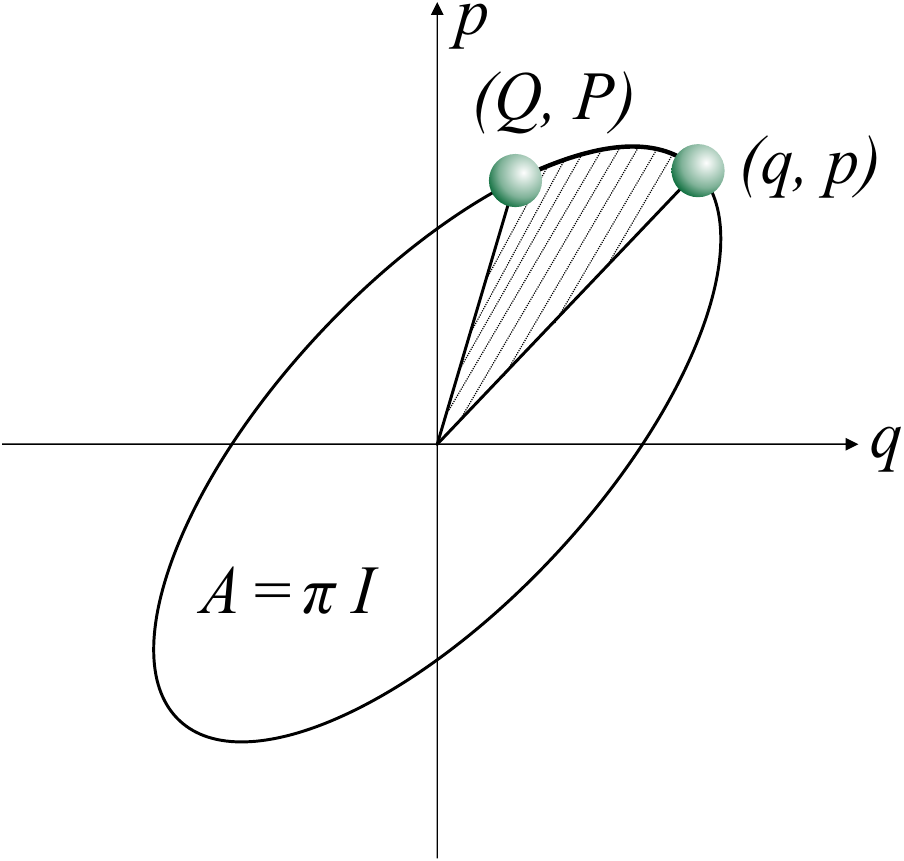}
\else
\includegraphics[height=.4\linewidth]{ellipse2.eps}
\fi
\end{center}
\caption{Visualization of the finite symmetry
transformation~(\ref{tdho-fini}) pertaining to the invariant
$I$ from Eq.~(\ref{tdinv}) of the time-dependent harmonic oscillator.}
\label{tdho-fig}
\end{figure}
The integral over the closed curve $C$ comprising the shaded
region $A_{\sigma}$ of Fig.~\ref{tdho-fig} measures the enclosed area
\begin{align*}
A_{\sigma}&=\onehalf\oint_{C}q\d p-p\d q
=\onehalf\int_{\phi}^{\phi+\sigma}\left(
q\dfrac{p}{\phi}-p\dfrac{q}{\phi}\right)\d\phi\\
&=\onehalf I\int_{\phi}^{\phi+\sigma}\left(
\cos^{2}\phi-\alpha_{e}\sin\phi\cos\phi+
\sin^{2}\phi+\alpha_{e}\sin\phi\cos\phi\right)\d\phi\\
&=\onehalf I\sigma.
\end{align*}
Note that the phase advance $\sigma$ does \emph{not} depict
the polar angle from vectors $(q,p)$ to $(Q,P)$.
Instead, $\sigma$ is proportional to the shaded area $A_{\sigma}$.
\subsubsection{\label{ex:tdho2}
Example: Symmetry of the time-dependent harmonic
oscillator at $\delta t\not=0$}
Replacing the quadratic $p$-dependence in the invariant~(\ref{tdinv})
of the time-dependent harmonic oscillator~(\ref{tdham}) according to
$$
\Hv=\onehalf p^{2}+\onehalf\omega^{2}(t)\,q^{2},
$$
we arrive at an \emph{equivalent} representation of the invariant
that now depends on the energy variable, $\Hv$
\begin{equation}\label{tdinv2}
I=2\beta_{e}(t)\,\Hv-\dot{\beta}_{e}(t)\,pq+\onehalf\ddot{\beta}_{e}(t)\,q^{2}.
\end{equation}
Of course, the function $\beta_{e}(t)$ must again satisfy
the second-order equation from Eq.~(\ref{tdinvcond}) in order
for $I$ to actually establish an invariant.
The particular infinitesimal rules for the corresponding
symmetry transformation from Eq.~(\ref{rules-infini2}) are
\begin{equation}\label{tdho-infini2}
\left.\begin{pmatrix}Q\\ P\end{pmatrix}\right|_{T}=
\left[\Eins+\AB_{\delta\epsilon}\right]
\left.\begin{pmatrix}q\\ p\end{pmatrix}\right|_{t},\quad
\AB_{\delta\epsilon}=\delta\epsilon\begin{pmatrix}-\dot{\beta}_{e}&0\\
-\ddot{\beta}_{e}&\dot{\beta}_{e}\end{pmatrix},\qquad
T=t-2\delta\epsilon\,\beta_{e}(t).
\end{equation}
As the coefficients of $\AB_{\delta\epsilon}$
do not explicitly depend on $\epsilon$,
we can set up the matrix exponential $\MB=\exp(\AB_{\delta\epsilon})$
according to the general scheme~(\ref{general-M}) in order
to finally derive the \emph{finite} transformation matrix
that corresponds to the infinitesimal mapping~(\ref{tdho-infini2}),
$$
\MB=\begin{pmatrix}\exp(-\delta\epsilon\,\dot{\beta}_{e})&0\\[\medskipamount]
-(\delta\epsilon\,\ddot{\beta}_{e}/\delta\epsilon\,\dot{\beta}_{e})
\sinh(\delta\epsilon\,\dot{\beta}_{e})&\;\;
\exp(\delta\epsilon\,\dot{\beta}_{e})\end{pmatrix},\qquad
\delta\epsilon=-\frac{\delta t}{2\beta_{e}}.
$$
Here, $\delta\epsilon$ still denotes an infinitesimal
$\epsilon$ interval.
The actual one-parameter symmetry transformation~(\ref{tdho-infini2})
is associated with a transformation of time $t\mapsto T$.
As the coefficients of $\AB_{\delta\epsilon}$ are time-derivatives
of the ellipse function $\beta_{e}(t)$ and thus generally depend on
time $t$, we must substitute $\delta\epsilon=-\delta t/2\beta_{e}(t)$
and \emph{integrate} all terms in $\MB$ that are proportional to
$\delta t$ over the finite interval $T-t$ that corresponds to a
\emph{finite} interval $\Delta\epsilon=\epsilon_{1}-\epsilon_{0}$,
\begin{align*}
m_{11}=\exp(-\delta\epsilon\,\dot{\beta}_{e})\quad\to\quad
m_{11}&=\exp\left(\int_{t}^{T}
\frac{\dot{\beta}_{e}(\tau)}{2\beta_{e}(\tau)}\d\tau\right)\\
&=\exp\left(\int_{t}^{T}\dfrac{}{\tau}\ln\sqrt{\beta_{e}}\,
\d\tau\right)=\sqrt{\frac{\beta_{e}(T)}{\beta_{e}(t)}}=\frac{1}{m_{22}}.
\end{align*}
With the identity
$\sinh\ln\,x=\left(x-x^{-1}\right)/2$,
the matrix element $m_{21}$ follows as
\begin{align*}
m_{21}=-\frac{\delta t\,\ddot{\beta}_{e}}{\delta t\,\dot{\beta}_{e}}
\sinh(\delta\epsilon\,\dot{\beta}_{e})\quad\to\quad
m_{21}&=-\frac{\dot{\beta}_{e}(T)-\dot{\beta}_{e}(t)}
{\beta_{e}(T)-\beta_{e}(t)}\sinh\ln\sqrt{\frac{\beta_{e}(t)}{\beta_{e}(T)}}\\
&=\frac{\dot{\beta}_{e}(T)-\dot{\beta}_{e}(t)}
{2\sqrt{\beta_{e}(T)\beta_{e}(t)}}.
\end{align*}
The \emph{finite} symmetry mapping $(q,p)_{t}\mapsto(Q,P)_{T}$
is thus finally obtained as
\begin{align}
\left.\begin{pmatrix}Q\\[\medskipamount] P\end{pmatrix}\right|_{T}&=
\frac{1}{\sqrt{\beta_{e}(T)\beta_{e}(t)}}
\begin{pmatrix}\beta_{e}(T)&0\\[\medskipamount]
\alpha_{e}(t)-\alpha_{e}(T)&\;\;
\beta_{e}(t)\end{pmatrix}
\left.\begin{pmatrix}q\\[\medskipamount] p\end{pmatrix}\right|_{t}
\label{tdho-fini2}\\
\Delta\sigma&=-2\Delta\epsilon=\int_{t}^{T}
\frac{\d\tau}{\beta_{e}(\tau)}.\nonumber
\end{align}
The symmetry mapping~(\ref{tdho-fini2}) is referred to
as the \emph{Floquet transformation}.
\subsubsection{\label{ex:rot-kepler}
Example: Rotational symmetry of the Kepler system}
The classical Kepler system is a two-body problem with the
mutual interaction following an inverse square force law.
In Cartesian coordinates, where no distinction between
covariant and contravariant coordinates is needed (all indexes lowered),
this system is described by a Hamiltonian
\begin{equation}\label{kepham}
H(\bq,\bp,t)=\onehalf p_{1}^{2}+\onehalf p_{2}^{2}+V(\bq,t)
\end{equation}
containing the interaction potential
\begin{equation}\label{keppot}
V(\bq,t)=-\frac{\mu(t)}{\sqrt{q_{1}^{2}+q_{2}^{2}}}=-\frac{\mu(t)}{r},
\end{equation}
with $\mu(t)=G\big[m_{1}(t)+m_{2}(t)\big]$ the possibly time-dependent
gravitational coupling strength that is induced by possibly
time-dependent masses $m_{1}$ and $m_{2}$ of the interacting bodies.
As the potential~(\ref{keppot}) spatially depends on $r$ only,
it is obviously invariant with respect to rotations in
configuration space $(q_{1},q_{2})$,
\begin{equation}\label{Kepler-rotation}
\begin{pmatrix}
Q_{1}\\ Q_{2}
\end{pmatrix}
=
\begin{pmatrix}
\hphantom{-}\cos\epsilon&\sin\epsilon\\
-\sin\epsilon&\cos\epsilon
\end{pmatrix}
\begin{pmatrix}
q_{1}\\ q_{2}
\end{pmatrix}
\end{equation}
where $\epsilon$ denotes the counterclockwise rotation angle.
This symmetry is not affected if we choose
$\epsilon\equiv\delta\epsilon$ to be very small.
We may then restrict ourselves in Eq.~(\ref{Kepler-rotation})
to first-order terms in $\delta\epsilon$ and insert the replacements
$\cos\delta\epsilon\approx 1$, $\sin\delta\epsilon\approx\delta\epsilon$.
This yields the infinitesimal transformation rules
\begin{equation}\label{Kepler-rotation-infini}
Q_{1}=q_{1}+\delta\epsilon\,q_{2},\qquad Q_{2}=q_{2}-\delta\epsilon\,q_{1}.
\end{equation}
This transformation can be regarded as being defined by a
generating function of the form of Eq.~(\ref{gen-infini}), namely
\begin{equation}\label{Kepler-rotation-gen}
\FC_{2}(q_{1},q_{2},P_{1},P_{2},t,\HCv)=-t\HCv+q_{1}P_{1}+q_{2}P_{2}+
\delta\epsilon\,(p_{1}q_{2}-p_{2}q_{1}).
\end{equation}
The transformation rules for the canonical momenta, energy, and
time emerge from the generating function~(\ref{Kepler-rotation-gen})
by applying the general canonical rules from Eqs.~(\ref{rules}),
$$
p_{1}=\pfrac{\FC_{2}}{q_{1}}=P_{1}-\delta\epsilon\,p_{2},\quad\!
p_{2}=\pfrac{\FC_{2}}{q_{2}}=P_{2}+\delta\epsilon\,p_{1},\quad\!
T=-\pfrac{\FC_{2}}{\HCv}=t,\quad\!\Hv=-\pfrac{\FC_{2}}{t}=\HCv.
$$
The rules from Eqs.~(\ref{Kepler-rotation-infini}) are indeed reproduced
as to first order in $\delta\epsilon$, we find the configuration space
transformation rules
$$
Q_{1}=\pfrac{\FC_{2}}{P_{1}}=q_{1}+\delta\epsilon\,q_{2},\qquad
Q_{2}=\pfrac{\FC_{2}}{P_{2}}=q_{2}-\delta\epsilon\,q_{1}.
$$
According to Eq.~(\ref{noetherinvariant4}), the expression
proportional to $\delta\epsilon$ in Eq.~(\ref{Kepler-rotation-gen})
must be a constant of motion in order for the infinitesimal
generating function $\FC_{2}$ to define a \emph{canonical}
transformation, hence to comply with the finite symmetry
transformation~(\ref{Kepler-rotation}) that preserves the
physical system.
Thus
$$
I=p_{1}q_{2}-p_{2}q_{1},\qquad\dfrac{I}{t}=0,
$$
which establishes the well-known conservation law of angular
momentum in --- possibly time-dependent --- central-force fields.
As the transformation rules~(\ref{Kepler-rotation}) only
depend on the parameter $\epsilon$ and \emph{not} on the
canonical variables, the transformation is referred
to as a \emph{global} symmetry transformation.

As with any generating function of a canonical transformation,
we can derive from Eq.~(\ref{Kepler-rotation-gen}) the rules
of both the configuration space coordinates and the
respective canonical momenta.
In matrix form, the infinitesimal rules for the momenta
can be rewritten as
$$
\begin{pmatrix}P_{1}\\ P_{2}\end{pmatrix}=
\left[\Eins+\AB_{\delta\epsilon}\right]
\begin{pmatrix}p_{1}\\ p_{2}\end{pmatrix},\qquad
\AB_{\delta\epsilon}=\delta\epsilon
\begin{pmatrix}\hphantom{-}0&1\\-1&0\end{pmatrix},
$$
with $\Eins$ denoting the $2\times 2$ unit matrix.
The corresponding \emph{finite} transformation is then
$$
\begin{pmatrix}P_{1}\\ P_{2}\end{pmatrix}=
\exp{(\AB_{\epsilon})}
\begin{pmatrix}p_{1}\\ p_{2}\end{pmatrix},\qquad
\exp{(\AB_{\epsilon})}=\begin{pmatrix}\hphantom{-}\cos\epsilon&
\sin\epsilon\\ -\sin\epsilon&\cos\epsilon\end{pmatrix},
$$
which coincides with the rules of the configuration
space variables from Eq.~(\ref{Kepler-rotation}).
This reflects the fact that the Hamiltonian~(\ref{kepham})
is equally invariant under rotations in momentum space.
\subsubsection{\label{ex:RL-kepler}
Example: Symmetry associated with the Runge-Lenz
invariant of the time-independent Kepler system}
As Noether's theorem associates the constants of motion of a dynamical
system with system symmetries, it can be applied in both directions.
In Sect.~\ref{ex:rot-kepler}, the constant of motion was determined
for a system symmetry that could be deduced directly from the form
of the Hamiltonian.
Conversely, if a constant of motion is known to exist,
then we can then derive the related system symmetry.
For the time-independent Kepler system~(\ref{kepham}), (\ref{keppot})
with $\mu=\text{const.}$, one component of the Runge-Lenz vector
is given by
\begin{equation}\label{Runge-Lenz1}
I_{1}=-q_{1}p_{2}^{2}+q_{2}p_{1}p_{2}+\mu\frac{q_{1}}{\sqrt{q_{1}^{2}+q_{2}^{2}}}.
\end{equation}
We easily convince ourselves that $I_{1}$ commutes with the
Hamiltonian $H$ from~(\ref{kepham}) with (\ref{keppot}).
Along the system's phase-space trajectory, we then have
$$
[I_{1},H]=0\quad\Longleftrightarrow\quad\dfrac{I_{1}}{t}=0.
$$
Using the invariant $I_{1}$ as the characteristic function $I$
in the generating function~(\ref{gen-infini}), the subsequent
transformation rules~(\ref{rules-infini}) then define the
corresponding infinitesimal symmetry transformation that
preserves the action functional~(\ref{canbed2a}).
The so obtained transformation is not particularly enlightening.
Yet, a better representation of the symmetry that is associated
with the Runge-Lenz invariant can be derived in the extended
Hamiltonian formalism.
In this context, we may express the invariant $I_{1}$ equivalently
as a function of $\bq$, $\bp$, and $\Hv$, with $\Hv$ being defined
as the \emph{value} of the Hamiltonian $H$ from Eq.~(\ref{kepham}),
$$
\Hv=\onehalf p_{1}^{2}+\onehalf p_{2}^{2}-
\frac{\mu}{\sqrt{q_{1}^{2}+q_{2}^{2}}}.
$$
The $\mu$-dependent term of the invariant $I_{1}$ can thus
be replaced by an $\Hv$-term according to
$$
\mu\frac{q_{1}}{\sqrt{q_{1}^{2}+q_{2}^{2}}}=
\onehalf q_{1}p_{1}^{2}+\onehalf q_{1}p_{2}^{2}-q_{1}\Hv,
$$
which yields an equivalent extended phase-space representation
of the Runge-Lenz invariant $I_{1}=I_{1}(\bq,\bp,\Hv)$
as a \emph{symmetric} quadratic form in the canonical momenta,
\begin{equation}\label{Runge-Lenz1a}
I_{1}=\onehalf q_{1}p_{1}^{2}+q_{2}p_{1}p_{2}-
\onehalf q_{1}p_{2}^{2}-q_{1}\Hv.
\end{equation}
As expected, the invariant $I_{1}$ commutes with the
Hamiltonian of the time-indepen\-dent Kepler system ($\mu=\text{const.})$
$$
{\left[I_{1},H\right]}_{\text{ext}}=p_{1}(H-\Hv)=0,
$$
hence establishes an invariant along the system's phase-space
trajectory as $H=\Hv$ by definition.
Due to the $\Hv$-dependence of the invariant $I_{1}$,
the corresponding symmetry transformation now includes a
transformation of time according to rules~(\ref{rules-infini2}).
Explicitly, the infinitesimal transformation rules are obtained as
\begin{align}
\delta p_{1}&=-\delta\epsilon\pfrac{I_{1}}{q_{1}}=\delta\epsilon\left(
\onehalf p_{2}^{2}-\onehalf p_{1}^{2}+\Hv\right)\qquad
\delta p_{2}=-\delta\epsilon\pfrac{I_{1}}{q_{2}}=
-\delta\epsilon\,p_{1}p_{2}\nonumber\\
\delta q_{1}&=\hphantom{-}\delta\epsilon\pfrac{I_{1}}{p_{1}}=
\delta\epsilon\left(q_{1}p_{1}+q_{2}p_{2}\right)\qquad\quad\:\,
\delta q_{2}=\hphantom{-}\delta\epsilon\pfrac{I_{1}}{p_{2}}=
\delta\epsilon\left(p_{1}q_{2}-p_{2}q_{1}\right)\nonumber\\
\delta\Hv&=\hphantom{-}\delta\epsilon\pfrac{I_{1}}{t}=0
\qquad\qquad\qquad\qquad\qquad\;\:
\delta t=-\delta\epsilon\pfrac{I_{1}}{\Hv}=\delta\epsilon\,q_{1}.
\label{rl-rules-infini}
\end{align}
The transformation rules for the new configuration space $Q_{1},Q_{2}$
variables depend \emph{linearly} on the original ones, $q_{1},q_{2}$.
We may thus rewrite the infinitesimal configuration space
transformation $Q_{i}=q_{i}+\delta q_{i},\; i=1,2$ in matrix form as
\begin{equation}\label{RL-rotation-infini}
{\left.\begin{pmatrix}Q_{1}\\ Q_{2}\end{pmatrix}
\right|}_{t+q_{1}\delta\epsilon}=
\left[\Eins+\AB_{\delta\epsilon}\right]
{\left.\begin{pmatrix}q_{1}\\ q_{2}\end{pmatrix}\right|}_{t},\qquad
\AB_{\delta\epsilon}(p_{1},p_{2})=
\delta\epsilon{\left.\begin{pmatrix}\hphantom{-}p_{1}&p_{2}\\
-p_{2}&p_{1}\end{pmatrix}\right|}_{t},
\end{equation}
with $\Eins$ denoting the $2\times 2$ unit matrix.
The form of the $2\times 2$ matrix $\AB_{\delta\epsilon}=(a_{ij})$ from
Eq.~(\ref{RL-rotation-infini}) with $a_{11}=a_{22}$ and $a_{12}=-a_{21}$
results from the particular representation~(\ref{Runge-Lenz1a})
of the Runge-Lenz invariant $I_{1}$.
With $\delta\epsilon$ still an \emph{infinitesimal} variation
of the parameter $\epsilon$, the transformation~(\ref{RL-rotation-infini})
can be expressed equivalently in terms of the matrix
exponential $\exp(\AB_{\delta\epsilon})$.
Then, the \emph{infinitesimal} symmetry transformation
then takes on the exceptionally simple form
\begin{equation}\label{RL-rotation}
{\left.\begin{pmatrix}Q_{1}\\ Q_{2}\end{pmatrix}
\right|}_{t+q_{1}\delta\epsilon}=
\exp{(p_{1}\,\delta\epsilon)}
\begin{pmatrix}\hphantom{-}\cos(p_{2}\,\delta\epsilon)&
\sin(p_{2}\,\delta\epsilon)\\
-\sin(p_{2}\,\delta\epsilon)&\cos(p_{2}\,\delta\epsilon)
\end{pmatrix}
{\left.\begin{pmatrix}q_{1}\\ q_{2}\end{pmatrix}\right|}_{t},
\end{equation}
The system symmetry that corresponds to the Runge-Lenz invariant
from Eq.~(\ref{Runge-Lenz1a}) is thus given by a \emph{local
scaled rotation} of the configuration space variables.
In contrast to the example of Sect.~\ref{ex:rot-kepler},
the transformation~(\ref{RL-rotation}) depends on the
actual coordinates $q_{1},p_{1},p_{2}$.
It is, therefore, referred to as a \emph{local} symmetry transformation.

Owing to the fact that the Hamiltonian~(\ref{kepham}) with
potential~(\ref{keppot}) is invariant under swappings
$q_{1}\leftrightarrow q_{2}$ \emph{and} $p_{1}\leftrightarrow p_{2}$,
the second component $I_{2}$ of the invariant Runge-Lenz
vector is obtained by flipping all indexes of $I_{1}$,
$$
I_{2}=\onehalf q_{2}p_{2}^{2}+q_{1}p_{1}p_{2}-
\onehalf q_{2}p_{1}^{2}-q_{2}\Hv.
$$
The infinitesimal transformation of the configuration space
coordinates follows as
$$
{\left.\begin{pmatrix}Q_{1}\\ Q_{2}\end{pmatrix}
\right|}_{t+q_{2}\delta\epsilon}=
\left[\Eins+\BB_{\delta\epsilon}\right]
{\left.\begin{pmatrix}q_{1}\\ q_{2}\end{pmatrix}\right|}_{t},\qquad
\BB_{\delta\epsilon}(p_{1},p_{2})=\delta\epsilon{\left.
\begin{pmatrix}p_{2}&-p_{1}\\
p_{1}&\hphantom{-}p_{2}\end{pmatrix}\right|}_{t}.
$$
Again, the transformation can be expressed equivalently in
terms of the matrix exponential $\exp(\BB_{\delta\epsilon})$,
where $\delta\epsilon$ denotes an infinitesimal shift of the
symmetry transformation's parameter
$$
{\left.\begin{pmatrix}Q_{1}\\ Q_{2}\end{pmatrix}
\right|}_{t+q_{2}\delta\epsilon}=
\exp{(p_{2}\,\delta\epsilon)}
\begin{pmatrix}\cos(p_{1}\,\delta\epsilon)&
-\sin(p_{1}\,\delta\epsilon)\\
\sin(p_{1}\,\delta\epsilon)&\hphantom{-}\cos(p_{1}\,\delta\epsilon)
\end{pmatrix}
{\left.\begin{pmatrix}q_{1}\\ q_{2}\end{pmatrix}\right|}_{t},
$$
\subsection{\label{ex:conv-noether}
Extended point transformations,
conventional Noether theorem}
The derivation of Noether's theorem in the context of the
Lagrangian formalism is restricted to \emph{extended point
transformations}, hence canonical transformations for which
the new space-time coordinates only depend on the old
space-time coordinates and \emph{not} on the set of old
momentum coordinates.
Yet, the extended canonical transformation approach allows
to describe more general possible symmetry mappings as the
rules~(\ref{rules-infini}) are \emph{not} restricted to
point transformations.
Consequently, equation~(\ref{gen-noether}) in conjunction with
the infinitesimal canonical mapping~(\ref{rules-infini})
represents a \emph{generalized formulation} of Noether's theorem.
In order to derive the \emph{conventional} Noether theorem
in the Hamiltonian description, we restrict ourselves to the
case of an infinitesimal point transformation, which is defined
by a generating function~(\ref{gen-infini}) with characteristic
function $I$ that is \emph{linear} in the momenta $p_{\nu}$
\begin{equation}\label{gen-infini2}
I(q^{\nu},p_{\nu})=-\sum_{\alpha=0}^{n}\eta^{\alpha}(q^{\nu})\,p_{\alpha}+
f(q^{\nu}),
\end{equation}
hence with functions $\eta^{\mu}=\eta^{\mu}(q^{\nu}),\,f=f(q^{\nu})$
that depend on the space-time coordinates only.
With this $I$, the transformation rules for space and time
coordinates follow as ($\mu,\nu=0,\ldots,n,\,i=1,\ldots,n$)
$$
\delta q^{\mu}=-\epsilon\eta^{\mu}(q^{\nu})\quad
\Leftrightarrow\quad\delta q^{i}=-\epsilon\eta^{i}(\bq,t),\quad
\delta t=-\epsilon\xi(\bq,t),\;\xi=\eta^{0}/c.
$$
The condition~(\ref{noetherinvariant}) for this transformation to preserve
the extended Hamiltonian $H_{\e}$, hence for the function~(\ref{gen-infini2})
to represent a conserved quantity along the system's evolution is
\begin{equation}\label{noetherinvariant2}
\sum_{\beta=0}^{n}\left[\eta^{\beta}\pfrac{H_{\e}}{q^{\beta}}+
\pfrac{H_{\e}}{p_{\beta}}\left(\pfrac{f}{q^{\beta}}-\sum_{\alpha=0}^{n}
p_{\alpha}\pfrac{\eta^{\alpha}}{q^{\beta}}\right)\right]=0.
\end{equation}
Distinguishing the canonical time and energy variables from
the canonical space and momentum coordinates, the Noether
function~(\ref{gen-infini2}) has the equivalent representation
\begin{equation}\label{gen-infini2a}
I(\bq,\bp,\Hv,t)=\xi(\bq,t)\,\Hv-\sum_{i=1}^{n}\eta^{i}(\bq,t)\,p_{i}+f(\bq,t),
\end{equation}
which represents a conserved quantity if
Eq.~(\ref{noetherinvariant3}) is satisfied.
In the last step, the energy variable $\Hv$ may be
replaced by the conventional Hamiltonian $H$.
We thus find the conventional Noether function in the Hamiltonian formulation
\begin{equation}\label{gen-infini2b}
I(\bq,\bp,t)=\xi(\bq,t)\,H-\sum_{i=1}^{n}\eta^{i}(\bq,t)\,p_{i}+f(\bq,t),
\end{equation}
which is an invariant provided that Eq.~(\ref{noetherinvariant3})
holds with $\partial I/\partial\Hv=0$.
Due to their different dependence on the canonical variables,
the Noether functions~(\ref{gen-infini2a}) and~(\ref{gen-infini2b})
yield different transformation rules from Eqs.~(\ref{rules-infini2}).
However, these rules are compatible as
\begin{equation}\label{gen-infini2c}
\delta\bar{p}_{i}=\delta p_{i}-\dfrac{p_{i}}{t}\delta t,\quad
\delta\bar{q}^{i}=\delta q^{i}-\dfrac{q^{i}}{t}\delta t,\quad
\delta\bar{\Hv}=\delta\Hv-\dfrac{H}{t}\delta t,\quad
\delta\bar{t}=0,
\end{equation}
if the barred quantities denote the variations derived from
Eq.~(\ref{gen-infini2b}) and the unbarred those derived
from Eq.~(\ref{gen-infini2a}).
As the function $I(\bq,\bp,t)$ does \emph{not depend} on the energy variable,
$\Hv$, the subsequent transformation rules are associated with an
\emph{identical} time transformation, $T=t,\;\delta\bar{t}=0$.
In contrast, $I(\bq,\bp,\Hv,t)$ from Eq.~(\ref{gen-infini2a})
accounts for an infinitesimal time shift transformation
$T=t-\epsilon\xi,\;\delta t=-\epsilon\xi$.
The connection of both equally valid sets of transformation
rules is given by Eqs.~(\ref{gen-infini2c}).

With these formulations, we are led to interpreting the
conventional Noether theorem in the reverse direction.
If we can find functions $f(\bq,t)$, $\xi(\bq,t)$, and $\eta^{i}(\bq,t)$
such that for a given conventional Hamiltonian $H$ the total time
derivative of $I$ vanishes, $\d I/\d t=0$, then the invariant $I$
in the forms of Eqs.~(\ref{gen-infini2a}) or~(\ref{gen-infini2b})
defines a corresponding extended canonical point transformation
according to Eqs.~(\ref{rules-infini2}).
\subsection{\label{sec:cq}Canonical quantization in the
extended Hamiltonian formalism}
The transition from classical dynamics to the corresponding
quantum description is most easily made in terms of the
``canonical quantization prescription.''
The quantum description of a dynamical system whose classical
limit is represented by a Hamiltonian $H$ is accordingly obtained
by reinterpreting our dynamical variables $q^{\mu}(s)$ and
$p_{\mu}(s)$ as \emph{operators} $\hat{q}^{\mu}(s)$ and
$\hat{p}_{\mu}(s)$ that act on a \emph{wave function} $\psi$.
In the configuration space representation,
the quantum mechanical operators are
\begin{equation}\label{oper-def}
\hat{q}^{\mu}=q^{\mu}\Eins,\qquad
\hat{p}_{\mu}=-i\hbar\pfrac{}{q^{\mu}},
\end{equation}
with $\Eins$ denoting the identity operator.
In the \emph{extended formalism}, an additional pair of
operators is given for the index $\mu=0$.
Because of $q^{0}\equiv ct$, $p_{0}\equiv-\Hv/c$,
these operators are expressed equivalently as
$$
\hat{t}=t\Eins,\qquad\hat{\Hv}=i\hbar\pfrac{}{t}.
$$
With $\Hv_{\e}$ denoting the \emph{value} of the extended
Hamiltonian $H_{\e}$, we encountered in Sect.~\ref{sec:caneq} another
additional pair of canonically conjugate variables, $(\Hv_{\e},s)$.
The corresponding operators are
$$
\hat{s}=s\Eins,\qquad\hat{\Hv}_{\e}=i\hbar\pfrac{}{s}.
$$
For explicitly $s$-dependent extended Hamiltonians $H_{\e}$
and wave functions $\psi(q^{\mu},s)$, the classical equation
$H_{\e}=\Hv_{\e}$ from Eq.~(\ref{p0-def1}) thus translates into
the equation of motion for the wave function $\psi(q^{\mu},s)$,
$$
\hat{H}_{\e}\,\psi=i\hbar\pfrac{\psi}{s}.
$$
This equation was postulated earlier by Feynman.\cite{feynman50}
The usual cases with no $s$-dependence of $H_{\e}$ and $\psi$
are then \emph{directly} obtained from the condition
$H_{\e}=0$ for the classical extended Hamiltonian~(\ref{hamid})
\begin{equation}\label{eq:gen-schroedinger}
\hat{H}_{\e}\,\psi(q^{\mu})=0.
\end{equation}
Equation~(\ref{eq:gen-schroedinger}) is the relativistic extension of the Schr\"odinger equation.

For the extended Hamiltonian of a point particle in an external
electromagnetic field from Eq.~(\ref{h1-em}), we immediately
find the Klein-Gordon equation, inserting Eqs.~(\ref{oper-def})
\begin{equation}\label{klein-gordon0}
\left[\left(i\hbar\pfrac{}{q^{\alpha}}+\frac{\zeta}{c}A_{\alpha}\right)
\left(i\hbar\pfrac{}{q_{\alpha}}+\frac{\zeta}{c}A^{\alpha}\right)+
m^{2}c^{2}\right]\psi(q^{\mu})=0.
\end{equation}
The non-relativistic limit is encountered by letting $s\to t$.
The corresponding extended Hamiltonian $H_{\e}=H-e=0$ from~(\ref{H1-triv}) with $H(\bq,\bp,t)$ a conventional
\emph{non-relativistic} Hamiltonian then yields the associated non-relativistic wave equation for $\psi(q^{\mu})\equiv\psi(\bq,t)$:
$$
\hat{H}\psi=i\hbar\pfrac{\psi}{t},
$$
which is referred to as the Schr\"odinger equation.
\subsection{\label{sec:kg}Path integral derivation of the
Klein-Gordon equation for a relativistic point particle
in an electromagnetic field}
Apart from the important additional rest energy term $-\onehalf mc^{2}$,
the extended Lagrangian~(\ref{lag1-em2}) for a relativistic
classical point particle in a external electromagnetic field
agrees with the Lagrangian proposed by Feynman\cite{feynman48}
on the basis of a formal reasoning.
We have seen that this Lagrangian $L_{\e}$ is actually \emph{not}
a mere formal construction, but has the physical meaning to
describe the \emph{same dynamics} as the corresponding
conventional Lorentz-invariant Lagrangian from Eq.~(\ref{lagr-em}).
As the extended Lagrangian~(\ref{lag1-em2}) is thus identified
as \emph{physically significant}, it can be concluded that the
path integral erected on this Lagrangian yields the correct quantum
description of a relativistic point particle in an external
electromagnetic field.

For an infinitesimal proper time step $\epsilon\equiv\Delta s$,
the action $S_{\e,\epsilon}$ for the extended
Lagrangian~(\ref{lag1-em2}) writes to first order in $\epsilon$
\begin{equation}\label{action1}
\qquad S_{\e,\epsilon}=\epsilon L_{\e}=\onehalf m\,\eta_{\alpha\beta}
\frac{(q^{\alpha}_{b}-q^{\alpha}_{a})
(q^{\beta}_{b}-q^{\beta}_{a})}{\epsilon}+
\frac{\zeta}{c}(q^{\alpha}_{b}-q^{\alpha}_{a})\,
A_{\alpha}(q^{\mu}_{c})-\onehalf mc^{2}\epsilon.
\end{equation}
The potentials $A_{\alpha}$ are to be taken at the space-time
location $q^{\mu}_{c}=(q^{\mu}_{b}+q^{\mu}_{a})/2$.
We insert this particular action function into Eq.~(\ref{trans-infini})
and perform a transformation of the integration variables $q^{\mu}_{a}$,
$$
q^{\mu}_{b}-q^{\mu}_{a}=\xi^{\mu}\quad\Rightarrow\quad
\d^{4}q_{a}=\d^{4}\xi.
$$
The integral~(\ref{trans-infini}) has now the equivalent representation
\begin{equation}\label{trans-infini2}
\psi(q^{\mu}_{b})=\frac{1}{M}\int\exp\left[
\frac{i}{\hbar}S_{\e,\epsilon}\right]\psi(q^{\mu}_{b}-\xi^{\mu})\,\d^{4}\xi,
\end{equation}
while the action $S_{\e,\epsilon}$ from Eq.~(\ref{action1}) takes on the form
$$
S_{\e,\epsilon}=\frac{m}{2}\eta_{\alpha\beta}\frac{\xi^{\alpha}\xi^{\beta}}{\epsilon}+
\frac{\zeta}{c}\xi^{\alpha}\!\left[A_{\alpha}(q^{\mu}_{b})\!-\!\onehalf
\xi^{\beta}\pfrac{A_{\alpha}(q^{\mu}_{b})}{q^{\beta}}\right]-
\epsilon\frac{mc^{2}}{2}.
$$
Here, we expressed the potentials $A_{\alpha}(q^{\mu}_{c})$ to first
order in terms of their values at $q^{\mu}_{b}$.
In the following, we skip the index ``$b$'' in the coordinate vector
as all $q^{\mu}$ refer to that particular space-time event
from this point of our derivation.

In order to match the quadratic terms in $S_{\e,\epsilon}$,
the wave function $\psi(q^{\mu}-\xi^{\mu})$ under the
integral~(\ref{trans-infini2}) must be expanded up to
second order in the $\xi^{\mu}$,
$$
\psi(q^{\mu}-\xi^{\mu})=\psi(q^{\mu})-\xi^{\alpha}
\pfrac{\psi(q^{\mu})}{q^{\alpha}}+\onehalf\xi^{\alpha}\xi^{\beta}
\pfrac{^{2}\psi(q^{\mu})}{q^{\alpha}\partial q^{\beta}}-\ldots
$$
The rest energy term in $S_{\e,\epsilon}$ depends only on $\epsilon$.
It can, therefore, be taken as a factor in front of the integral
and expanded up to first order in $\epsilon$.
The total expression~(\ref{trans-infini2}) for the transition
of the wave function $\psi$ thus follows as
\begin{align}
\psi=\frac{1}{M}
\left(1-\epsilon\frac{imc^{2}}{2\hbar}\right)
&\int_{-\infty}^{\infty}\exp\left\{\frac{i}{\hbar\epsilon}\left[
\frac{m}{2}\eta_{\alpha\beta}\xi^{\alpha}\xi^{\beta}+
\frac{\zeta\epsilon}{c}A_{\alpha}\xi^{\alpha}-\frac{\zeta\epsilon}{2c}
\pfrac{A_{\alpha}}{q^{\beta}}\xi^{\alpha}\xi^{\beta}\right]\right\}
\nonumber\\
&\times\left[\psi-\xi^{\alpha}
\pfrac{\psi}{q^{\alpha}}+\onehalf\xi^{\alpha}\xi^{\beta}
\pfrac{^{2}\psi}{q^{\alpha}\partial q^{\beta}}\right]\d^{4}\xi.
\label{trans-infini3}
\end{align}
Prior to actually calculating the Gaussian type integrals,
we may simplify the integrand in~(\ref{trans-infini3})
by taking into account that the third term in the exponential
function is of order of $\epsilon$ smaller than the first one.
We may thus factor out this term and expand it up to first order in $\epsilon$
$$
\exp\left[-\frac{i\zeta\epsilon}{2\hbar c}\pfrac{A_{\alpha}}{q^{\beta}}
\xi^{\alpha}\xi^{\beta}\right]=1-\frac{i\zeta\epsilon}{2\hbar c}
\pfrac{A_{\alpha}}{q^{\beta}}\xi^{\alpha}\xi^{\beta}+\ldots
$$
Omitting terms of higher order than quadratic in the $\xi^{\mu}$,
the integral becomes
\begin{align*}
\psi=\frac{1}{M}
\left(1-\epsilon\frac{imc^{2}}{2\hbar}\right)
&\int_{-\infty}^{\infty}\exp\left\{\frac{i}{\hbar}\left[
\frac{m}{2\epsilon}\eta_{\alpha\beta}\xi^{\alpha}\xi^{\beta}+
\frac{\zeta}{c}A_{\alpha}\xi^{\alpha}\right]\right\}\\
&\!\!\times\left[\psi-\xi^{\alpha}
\pfrac{\psi}{q^{\alpha}}+\onehalf\xi^{\alpha}\xi^{\beta}\left(
\pfrac{^{2}\psi}{q^{\alpha}\partial q^{\beta}}-\frac{i\zeta}{\hbar c}
\pfrac{A_{\alpha}}{q^{\beta}}\,\psi\right)\right]\d^{4}\xi.
\end{align*}
The integral over the entire space-time can now be solved
analytically to yield
\begin{align*}
\psi&=\frac{1}{M}{\left(
\frac{2\pi\hbar\epsilon}{im}\right)}^{2}
\left(1-\epsilon\frac{imc^{2}}{2\hbar}\right)
\exp\left\{-\epsilon\frac{i\zeta^{2}}{2\hbar mc^{2}}
A^{\alpha}A_{\alpha}\right\}\\
&\quad\times\left[\psi+\epsilon\frac{\zeta}{mc}A^{\alpha}
\pfrac{\psi}{q^{\alpha}}+\frac{\epsilon}{2}\left(
\pfrac{^{2}\psi}{q^{\alpha}\partial q^{\beta}}
-\frac{i\zeta}{\hbar c}\pfrac{A_{\alpha}}{q^{\beta}}\psi\right)
\left(\frac{\epsilon\zeta^{2}}{m^{2}c^{2}}A^{\alpha}A^{\beta}+
\frac{i\hbar}{m}\eta^{\alpha\beta}\right)\right].
\end{align*}
We may omit the term quadratic in $\epsilon$ that is contained
in the rightmost factor and finally expand the exponential function
up to first order in $\epsilon$
\begin{align}
\psi&=\frac{1}{M}{\left(\frac{2\pi\hbar\epsilon}{im}\right)}^{2}
\left(1-\epsilon\frac{imc^{2}}{2\hbar}\right)
\left(1-\epsilon\frac{i\zeta^{2}}{2\hbar mc^{2}}
A^{\alpha}A_{\alpha}\right)\nonumber\\
&\quad\times\left[\psi+\epsilon\frac{\zeta}{mc}A^{\alpha}
\pfrac{\psi}{q^{\alpha}}+\epsilon\frac{i\hbar}{2m}\left(
\pfrac{^{2}\psi}{q^{\alpha}\partial q_{\alpha}}
-\frac{i\zeta}{\hbar c}\pfrac{A^{\alpha}}{q^{\alpha}}\psi\right)\right].
\label{kg1}
\end{align}
The normalization factor $M$ is now obvious.
As the equation must hold to zero order in
$\epsilon$, we directly conclude that
$M={\left(2\pi\hbar\epsilon/im\right)}^{2}$.
This means, furthermore, that the sum over all terms
proportional to $\epsilon$ must vanish.
The five terms in~(\ref{kg1}) that are linear in $\epsilon$
thus establish the equation
$$
\frac{m^{2}c^{2}}{\hbar^{2}}\psi=
\pfrac{^{2}\psi}{q^{\alpha}\partial q_{\alpha}}
-\frac{\zeta^{2}A^{\alpha}A_{\alpha}}{\hbar^{2}c^{2}}
\psi+\frac{2\zeta A^{\alpha}}{i\hbar c}
\pfrac{\psi}{q^{\alpha}}+\frac{\zeta}{i\hbar c}
\pfrac{A^{\alpha}}{q^{\alpha}}\psi.
$$
This equation has the equivalent product form
\begin{equation}\label{klein-gordon}
\left(\pfrac{}{q^{\alpha}}-\frac{i\zeta}{\hbar c}A_{\alpha}\right)
\left(\pfrac{}{q_{\alpha}}-\frac{i\zeta}{\hbar c}A^{\alpha}\right)
\psi={\left(\frac{mc}{\hbar}\right)}^{2}\psi,
\end{equation}
which constitutes exactly the Klein-Gordon equation
for our metric $\eta_{\mu\nu}$.
It coincides with the wave equation~(\ref{klein-gordon0})
that emerged from the canonical quantization formalism.

We remark that Feynman\cite{feynman50} went the
procedure developed here in the opposite direction.
He started with the Klein-Gordon equation and deduced
from analogies with the non-relativistic case a
classical Lagrangian similar to that of Eq.~(\ref{lag1-em2}),
but without its rest energy term $-\onehalf mc^{2}$.
The obtained Lagrangian was \emph{not} identified as
\emph{physically significant}, i.e., as exactly the extended
Lagrangian $L_{\e}$ that describes the corresponding classical
system, but rated as ``purely formal.''\cite{feynman48}
\subsection{\label{sec:prop}Space-time kernel
for the free relativistic point particle}
The hypersurface condition~(\ref{constraint-lag}) is to be
disregarded setting up the parameterized kernel~(\ref{kernel-para})
as virtual particles are to be included.
The components of the extended free-particle Lagrangian~(\ref{lag1-fp})
can then be treated as \emph{independent}.
The corresponding action functional $S$ from Eq.~(\ref{principle1})
thus splits into a sum of independent action functionals,
\begin{equation}\label{actint-fp}
S_{\e}[q^{\nu}(s)]=\onehalf m\int_{s_{a}}^{s_{b}}\left(
\dfrac{q^{\alpha}}{s}\dfrac{q_{\alpha}}{s}-c^{2}\right)\d s=
\sum_{\alpha}S[q^{\alpha}(s)].
\end{equation}
Hence, the parameterized space-time kernel~(\ref{kernel-para})
separates into a product of path integrals.
For the free particle, the individual path integrals
can be solved analyti\-cally.\cite{feynman,kleinert}
Expressed in terms of $s$ as the independent variable,
the result for one degree of freedom $q^{k}$ is
\begin{equation}\label{freekernel-1d}
K_{s}(q^{k}_b,q^{k}_a)=\sqrt{\frac{m}{2\pi i\hbar(s_{b}-s_{a})}}\exp
\left[\frac{im}{2\hbar}\frac{{(q^{k}_{b}-q^{k}_{a})}^{2}}{s_{b}-s_{a}}\right].
\end{equation}
The total parameterized space-time kernel $K_{\sigma}(b,a)$
is then obtained for $S_{\e}$ from Eq.~(\ref{actint-fp}) as
$$
K_{s}(b,a)=-\frac{m^{2}c}{4\pi^{2}\hbar^{2}{(s_{b}-s_{a})}^{2}}
\exp\left\{\frac{im}{2\hbar}\left[\frac{(q^{\alpha}_{b}-q^{\alpha}_{a})(q_{\alpha,b}-
q_{\alpha,a})}{s_{b}-s_{a}}-c^{2}(s_{b}-s_{a})\right]\right\}.
$$
The term proportional to \mbox{$(s_{b}-s_{a})$} in the exponential function
originates from the rest energy term $-\onehalf mc^{2}$ in the
extended Lagrangian~(\ref{lag1-fp}) and, correspondingly,
in the action integral~(\ref{actint-fp}).
The integration over the parameter variable $s$ is worked out by means of a Wick rotation.
The parameter interval is then $\sigma=i(s_{b}-s_{a})$.
With $\tau$ defined by
$$
\tau^{2}=\frac{(q^{\alpha}_{b}-q^{\alpha}_{a})(q_{\alpha,b}-q_{\alpha,a})}{c^{2}},
$$
the parameterized space-time kernel $K_{\sigma}(b,a)$ takes on the equivalent form
$$
K_{\sigma}(b,a)=\frac{m^{2}c}{4\pi^{2}\hbar^{2}}\,\sigma^{-2}
\exp\left[-\frac{mc^{2}}{2\hbar}\left(
\frac{\tau^{2}}{\sigma}+\sigma\right)\right].
$$
According to Eq.~(\ref{kernel-gen}), the space-time propagator
$K(b,a)$ for a free relativistic wave packet is finally acquired
by integrating $K_{\sigma}(b,a)$ over all parameter intervals $\sigma$
\begin{equation}\label{kernel0-fp}
K(b,a)=\frac{m^{2}c}{4\pi^{2}\hbar^{2}}\int_{0}^{\infty}
\sigma^{-2}\exp\left[-\frac{mc^{2}}{2\hbar}\left(
\frac{\tau^{2}}{\sigma}+\sigma\right)\right]\d\sigma.
\end{equation}
The integral is proportional to the integral representation of the
Bessel function $K_1$ of second kind and order one\cite{magnus},
that is also referred to as MacDonald function,
\begin{equation}\label{intrep}
\int_{0}^{\infty}\sigma^{-2}\exp\left[
-\frac{M}{2}\left(\frac{\tau^{2}}{\sigma}+
\sigma\right)\right]\d\sigma=\frac{2}{\tau}K_{1}(M\tau),\qquad M=\frac{mc^2}{\hbar}.
\end{equation}
For our metric $\eta_{\mu\nu}=\mathrm{diag}(-1,1,1,1)$,
a positive $\tau^{2}$ represents a \emph{space-like} connection of the events $a$ and $b$.
The kernel $K(b,a)$ from Eq.~(\ref{kernel0-fp}) is then given by
\begin{equation}\label{kernel-expl}
K(b,a)=\frac{m^{2}c^2}{2\pi^{2}\hbar^{2}}\,\frac{1}{|q_b-q_a|}K_{1}\left(\frac{mc}{\hbar}|q_b-q_a|\right),\qquad\tau^2>0.
\end{equation}
If $\tau^{2}$ is negative, one encounters a \emph{time-like} connection of the events $a$ and $b$.
The kernel $K(b,a)$ is then expressed in terms of the Hankel function $H_1^{(1)}(x)=-\frac{2}{\pi}K_1(i x)$ as:
\begin{equation}\label{kernel-expl-hankel}
K(b,a)=\frac{im^{2}c^2}{4\pi\hbar^{2}}\,\frac{1}{|q_b-q_a|}H_{1}^{(1)}\left(\frac{mc}{\hbar}|q_b-q_a|\right),\qquad\tau^2<0.
\end{equation}
We may convince ourselves by direct substitution that the kernels~(\ref{kernel-expl})
and~(\ref{kernel-expl-hankel}) satisfy the zero-potential case \mbox{($A_{\mu}=0$)}
of the Klein-Gordon equation~(\ref{klein-gordon}):
\begin{equation*}
\ppfrac{}{q^{\alpha}}{q_{\alpha}}K(b,a)=\pm\frac{m^2c^2}{\hbar^2}\,K(b,a).
\end{equation*}
As a consequence, so does a free-particle wave function
$\psi(\bq,t)$ if its space-time propagation is calculated
according to Eq.~(\ref{wave-evol}).

In order to determine the non-relativistic limit $c\to\infty$ of Eq.~(\ref{kernel-expl}),
we consider the asymptotic behavior of $\tau$ and the Bessel function $K_{1}$:
\begin{align*}
\tau=\sqrt{-{(t_{b}-t_{a})}^{2}+{(\bq_{b}-\bq_{a})}^{2}/c^{2}}
&\quad\stackrel{c\to\infty}{=}\quad i(t_{b}-t_{a})\\
\frac{1}{\tau}K_1(M\tau)&\quad\stackrel{c\to\infty}{=}\quad\sqrt{\frac{\pi}{2M\tau^3}}\,\exp(-M\tau)\\
\exp\left(-\frac{mc^{2}}{\hbar}\tau\right)
&\quad\stackrel{c\to\infty}{=}\quad\exp\left[\frac{im}{2\hbar}
\frac{{(\bq_{b}-\bq_{a})}^{2}}{t_{b}-t_{a}}\right].
\end{align*}
The nonrelativistic kernel $K(b,a)$ the kernel for three spatial degrees of freedom becomes
$$
K_{\bq}(b,a)={\left[\frac{m}{2\pi i\hbar(t_{b}-t_{a})}\right]}^{3/2}\,
\exp\left[\frac{im}{2\hbar}\frac{{(\bq_{b}-\bq_{a})}^{2}}{t_{b}-t_{a}}\right].
$$
This kernel generalizes the one-dimensional case
(Eq.~{\ref{freekernel-1d}})
and satisfies again the Schr\"odinger equation\cite{feynman,kleinert}.
\section{Conclusions}
Starting from the space-time formulation of the action principle,
we have demonstrated that the Lagrangian as well as the Hamiltonian
description of classical dynamics can consistently be
reformulated in order to be compatible with special relativity.
In the emerging \emph{extended} version of the Hamilton-Lagrange
formalism, the dynamics is described as a motion on a hypersurface
within an \emph{extended} phase space.
With the specific correlations of extended Lagrangian
$L_{\e}$ and extended Hamiltonian $H_{\e}$ to their
conventional counterparts $L$ and $H$ given in this paper,
the extended formalism retains the \emph{form} of the
long-established conventional Hamilton-Lagrange formalism.
The extended Hamilton-Lagrange formalism thus provides an
\emph{equivalent physical description} of dynamical systems
that is particularly appropriate for special relativity.

The physical significance of the Lorentz invariant extended
Hamiltonian $H_{\e}$ of a point particle in an external electromagnetic
field was demonstrated by showing that the subsequent \emph{extended}
set of canonical equations, in conjunction with the condition $H_{\e}=0$,
is \emph{equivalent} to the set of canonical equations that follows
from the well-known conventional Hamiltonian $H$ for this system.
It was shown that the condition $H_{\e}=0$ is automatically satisfied
on the system path that is defined by the solution of the canonical equations.
For this reason, the hypersurface condition $H_{\e}=0$ actually
does \emph{not} represent a constraint for the system.
The corresponding non-homogeneous extended Lagrangian $L_{\e}$ was shown to be \emph{quadratic}
in its velocity terms, hence similar in its \emph{form} with the
conventional Lagrangian $L$ that describes the non-relativistic limit.
This makes the extended formalism particularly suited for
analytical approaches that depend on the Lagrangian to be quadratic
in the velocities --- like Feynman's path integral formalism.
Devising the ``quantum version'' of the action principle, one
of Feynman's achievements was to derive --- by means of his path
integral approach to quantum physics --- the Schr\"odinger equation
as the quantum description of a physical system whose classical
limit is described by the non-relativistic Lagrangian $L$ for
a point particle in an external potential.
This is generally regarded as the \emph{proof of principle}
for the path integral formalism.

Similar to the extension of the conventional Hamilton-Lagrange
formalism in the realm of classical physics, the general form
of the relativistic extension of Feynman's path integral
approach is obtained by consistently treating space and time
variables on equal footing.
We have shown that the hypersurface condition from the classical extended
formalism appears in the context of the extended path integral
formalism as an \emph{additional uncertainty relation}.

On the basis of the extended Lagrangian $L_{\e}$ of a classical
relativistic point particle in an external electromagnetic field,
we could derive the Klein-Gordon equation as the corresponding
quantum description by means of the space-time version
of the path integral formalism.
Correspondingly, we can regard the emerging of the
Klein-Gordon equation as the proof of principle of the
\emph{relativistic generalization} of Feynman's path
integral approach that is based on Lorentz invariant
\emph{extended Lagrangians} $L_{\e}$ in conjunction
with the additional \emph{uncertainty relation}.
\section*{Acknowledgment}
The author is indebted to Prof.~Dr.~Walter Greiner from the
\emph{Frankfurt Institute of Advanced Studies} (FIAS) for
his critical comments and encouragement.

\end{document}